\documentclass[a4paper,10pt]{article}
\usepackage{amsmath,amscd}
\usepackage{amssymb}
\usepackage{amsmath}
\usepackage{amsfonts}
\usepackage{pxfonts}
\usepackage{color}
\def\bbbn{{\mathbb N}}
\def\bbbc{{\mathbb C}}

\def\bbbz{{\mathbb Z}}
\def\bbbr{{\mathbb R}}
\def\bbbd{{\mathbb D}}

\def\cp1{{\mathbb C\mathbb P}^1}
\def\cH{{\cal H}}

\def\cE{{\cal S}}
\def\cR{{\cal R}}
\def\cA{{\cal A}}
\def\cK{{\cal K}}

\def\cR{{\cal R}}

\def\cC{{\cal C}}

\def\cF{{\cal F}}
\def\acS{{\, \xrightarrow{\cS}\, }}

\newcommand\J{{\mathcal I}}
\newcommand\cS{{\mathcal S}}

\newtheorem{Def}{Definition}

\newtheorem{Ex}{Example}

\def \bG {{\bf G}}
\def \bF {{\bf F}}
\def \bK {{\bf K}}
\def \bu {{\bf u}}
\def \bv {{\bf v}}
\def \ba {{\bf a}}
\def \bg {{\bf g}}
\def \bI {{\bf I}}
\begin{document}

\title{Darboux transformations and Recursion operators for differential--difference equations}
\author{Farbod Khanizadeh$^{\dagger}$, Alexander V. Mikhailov$^{\star}$ and Jing Ping Wang$ ^\dagger $
\\
$\dagger$ School of Mathematics, Statistics \& Actuarial Science, University of Kent, UK \\
$\star$ Applied Mathematics Department, University of Leeds, UK
}
\date{}
\maketitle
\begin{abstract}
In this paper we review two concepts directly related to the Lax representations: Darboux transformations and Recursion operators
for integrable systems. We then present an extensive list of integrable differential-difference equations
together with their Hamiltonian structures, recursion operators, nontrivial generalised symmetries and Darboux-Lax representations.
The new results include multi-Hamiltonian structures and recursion operators for integrable Volterra type equations, integrable 
discretization of derivative nonlinear Schr\"odinger equations such as 
the Kaup-Newell lattice, the Chen-Lee-Liu lattice and the Ablowitz-Ramani-Segur (Gerdjikov-Ivanov) lattice.
We also compute the weakly nonlocal inverse recursion operators.
\end{abstract}

\section{Introduction}
The aim of this paper is to give a comprehensive account of multi-Hamiltonian structures, 
recursion operators and Darboux-Lax representations for a wide class of integrable differential-difference 
equations. Some of these results are well known, but scattered in literature.
In many cases we have completed the picture providing explicit expressions for
Hamiltonian, symplectic, recursion operators and Darboux-Lax representations. The Lax representations of nonlinear differential and difference equations play a central role in the theory of 
integrable systems. 
It enables one to apply the inverse transform method for construction  of exact solutions and study the 
asymptotic of the initial value problem.
 It also enables one to construct  a recursion operator which generates a infinite hierarchies of 
symmetries and conservation laws. Currently for a given equation there is not any general method to find its Lax 
representation. The most successful approach is the Wahlquist--Estabrook prolongation procedure \cite{WE73}. Recently, 
Mikhailov and his coauthors have tackled this problem from a different angle \cite{LM04,LM05, sara, bury}. 
They studied possible reductions of a general Lax representation using the reduction group approach 
\cite{mik79, mik80, mik81} and further leading to a classification of the Lax representations and corresponding
integrable equations.

The concept of Darboux transformations is originated from the classical differential 
geometry \cite{mr1908706}. Application of Darboux transformations to the corresponding Lax representation
leads to B\"acklund transformations and to a generation of new exact solutions for the integrable system.  
B\"acklund transformations can be regarded as an integrable system of differential-difference equations on their own rights.
These differential-difference equations play role of infinitesimal  symmetries for the integrable partial difference equations which 
can be obtained from the condition of the Bianchi commutativity of the Darboux transformations.

Evolutionary differential-difference  systems are the main object of our study. Our notations are standard.
We illustrate them on the well-known example of the Volterra equation \cite{Volterra} 
(cf. Section \ref{vol} for more algebraic properties of this equation), which we can write in the form
\begin{equation}\label{vol0}
u_{t}=u(u_{1}-u_{-1}).
\end{equation}
Here the dependent variable $u$ is assumed to be a function $u(n,t)$ of a lattice variable $n\in \bbbz$ and a continuous variable $t$ 
and we denote
\begin{eqnarray*}
 u_t=\partial_t u(n,t), \quad u_j=u(n+j,t).
\end{eqnarray*}
We shall omit the subscript index zero and use $u$ instead of $u_0$.
The Volterra equation (\ref{vol0}) encodes an infinite sequence of differential equations 
$$\partial_t u(n,t)=u(n,t)(u(n+1,t)-u(n-1,t)),\quad n\in\bbbz.$$ 

Equation (\ref{vol0}) possesses an infinite hierarchy of symmetries, or in other words, it is
compatible  with an infinite sequence of evolutionary equations of the form
\[ u_{t_m}=K_m(u_{m},u_{m-1},\ldots , u_{1-m},u_{-m}), \qquad m\in \bbbn \]
where $t_1=t, \ K_1=u(u_{1}-u_{-1})$,
\[
 K_2=u \left(u_{1} u_{2}+u_{1}^2+u u_1-u u_{-1}-u_{-1}^2- u_{-1} u_{-2}\right)
\]
 and $K_m$ are certain polynomials of variables $u_{m},u_{m-1},\ldots , u_{1-m},u_{-m} $.
The compatibility means  $ \partial_{t_m}(\partial_t u)= \partial_{t}(\partial_{t_m} u)$ or
the vanishing of the Lie bracket $[K_1,K_m]$, which is defined as
\[
[K_1,K_m]\coloneqq K_{m\star}(K_1)-K_{1\star}(K_m)=0
\]
where $K_{m\star}$ denotes the Fr\'echet derivative
\[
 K_{m\star}=\sum_{p=-m}^m \frac{\partial K_m}{\partial u_p}\cS^p
\]
and $\cS$ denotes the shift operator, such that $\cS (u_k)=u_{k+j}$ and for any function $f(u_{k_1},\ldots,u_{k_2})$ we have
$\cS^j (f(u_{k_1},\ldots,u_{k_2}))=f(u_{k_1+j},\ldots,u_{k_2+j})$.

Symmetries of the Volterra equation can be generated  by the recursion operator $\cR$
\begin{equation}\label{recvol0}
\cR=u\cE+u+u_1+u \cE^{-1}+u(u_1-u_{-1})(\cE-1)^{-1}\frac{1}{u} ,
\end{equation}
namely
$$K_{m+1}=\cR^m(K_1)$$

The vector fields corresponding to the Volterra equation and its symmetries are difference polynomials, i.e. elements of the 
difference polynomial ring $R=[\bbbc, u, \cS]$ which is a multi-variate ring  over the field of complex numbers 
$\bbbc$ with infinite number of indeterminates $u_k,\ k\in\bbbz$ and equipped with 
the automorphism $\cS$. The corresponding field of fractions $\cF=(\bbbc, u, \cS)$ is a difference field of rational 
functions of indeterminates $u_k,\ k\in\bbbz$ over $\bbbc$, which inherits the automorphism $\cS$. 
 Difference operators are defined as finite sums of the form $\sum a_k\cS^k$, where $a_k\in \cF$ 
(the above defined Fr\'echet derivative $K_{m\star}$ is an example of a difference operator). Difference operators 
act naturally on elements of $\cF$.
Similar to the differential case the elements of the ring $R$, field $\cF$ and difference operators are called local polynomials, local functions and local operators 
respectively. 

The recursion operator $\cR$ given by (\ref{recvol0}) is not a local operator. It contains the local part and the term 
$u(u_1-u_{-1})(\cE-1)^{-1}u^{-1}$.  Action of $\cR$ can be defined on those elements of $\cF$ which belong to the the image of 
the difference operator $u(\cS-1)$. We can directly check that $K_1,K_2\in \mbox{Im }(u(\cS-1))$. Similar to \cite{mwx2} it can be shown that all
 $K_{m+1}=\cR^m(K_1), \ m\in\bbbn$ are difference polynomials. The action of $\cR$ is not uniquely defined on the elements of 
$\mbox{Im }(u(\cS-1))$. Indeed, the base field $\bbbc$ is the kernel space of the difference operator $u(\cS-1)$ and thus 
for $a=u(\cS-1)(b),\ b\in\cF$ we have $(\cE-1)^{-1}(u^{-1}a)=b+\alpha$, where $\alpha\in\bbbc $ is an arbitrary constant. 
Obviously, acting by the recursion operator $\cR$ defined by (\ref{recvol0}) on $K_1$ we get
\[ \cR(u(u_1-u_{-1}))=u \left(u_{1} u_{2}+u_{1}^2+u u_1-u u_{-1}-u_{-1}^2- u_{-1} u_{-2}\right)+\alpha u(u_1-u_{-1}).
\]
Action of the recursion operator $\cR$ is well defined on the sequence of 
the quotient linear spaces $$\cK_1=\mbox{Span}_\bbbc (K_1),\quad 
 \cK_m=\mbox{Span}_\bbbc (K_1,K_2,\ldots,K_m)/  \mbox{Span}_\bbbc (K_1,K_2,\ldots,K_{m-1})$$ and in what follows 
describing the result of the action of a recursion operator on a symmetry we shall give one representative 
from the corresponding co-set.

The recursion operator  $\cR$ given by (\ref{recvol0}) is a pseudo-difference operator. It can be represented in the form 
 $\cR=B A^{-1}$, where $A$ and $B$ are difference operators. For example one can take $A=\cH_1,B=\cH_2$, where    $\cH_1$ and $\cH_2$
are two Hamiltonian operators for the Volterra equation (see Section 4.1).  Notice that the pseudo-difference operator $\cR$
is a sum of a local (difference) operator and a  nonlocal term of the form $ P (\cE-1)^{-1} Q$.  We say that a pseudo-difference 
operator is weakly nonlocal if it can be represented in the form 
\[
 A+\sum_i P_i(\cS-1)^{-1} Q_i
\]
where $A,P_i$ and $Q_i$ are difference operators and the sum is finite (a similar terminology has been first introduced in the study 
of pseudo-differential Hamiltonian operators \cite{MaN01}). Thus the recursion operator  $\cR$ given by (\ref{recvol0}) 
is a weakly nonlocal pseudo-difference operator. Moreover, it is easy to show that $\cR^m,\ m\in \bbbn$ is a pseudo-difference operator.  
In the majority of cases studied in the present paper the recursion operators are
weakly nonlocal. Exceptions include the recursion operator for the Narita-Itoh-Bogoyavlensky lattice \cite{wang12} (see Section 4.5).

We notice that surprisingly, in the case of multi-component  systems of integrable difference equations with weakly nonlocal recursion operators 
often the inverse recursion operator is also weakly nonlocal. It enables us to generate infinitely many 
local  symmetries corresponding to the inverse flows. For example the Heisenberg ferromagnet lattice \cite{sklyanin} 
(cf. Section \ref{heisen} for more algebraic properties of this equation) 
\begin{eqnarray*}
\left\{ {\begin{array}{l} u_t= (u-v) (u-u_1) (u_1-v)^{-1}\\
   v_t= (u-v) (v_{-1}-v) (u-v_{-1})^{-1}  \end{array} } \right.
 \end{eqnarray*}
possesses a recursion operator
\begin{eqnarray*}
&& \cR=\left( {\begin{array}{cc}
 \frac{(u-v)^2}{(u_1-v)^{2}} \cS -\frac{2 (u-u_1) (v-v_{-1})}{(u-v_{-1}) (u_1-v)}& -\frac{(u-u_1)^2}{ (u_1-v)^{2}}  \\
 \frac{(v-v_{-1})^2}{ (u-v_{-1})^{2}} & \frac{(u-v)^2}{ (u-v_{-1})^{2}} \cS^{-1}
 \end{array} } \right)+2 K^{(1)} (\cS-1)^{-1} Q^{(1)},
\end{eqnarray*}
where 
\[ 
K^{(1)}= \left(\begin{array}{c}\frac{(u-v) (u-u_1) }{u_1-v}\\
  \frac{ (u-v) (v_{-1}-v)}{ u-v_{-1}}  \end{array} \right),\quad  
Q^{(1)}=\left(\begin{array}{cc} \frac{v-v_{-1}}{(u-v) (u-v_{-1})}, & \frac{u-u_1}{(u-v) (u_1-v)}  \end{array}\right)
\] 
The operator $\cR$ is weakly nonlocal and it has a weakly nonlocal inverse
\begin{eqnarray*}
&& \cR^{-1}=\left( {\begin{array}{cc}
 \frac{(u-v)^2}{(u_{-1}-v)^{2}} \cS^{-1}& \frac{(u-u_{-1})^2}{ (u_{-1}-v)^{2}}  \\
 -\frac{(v-v_{1})^2}{ (u-v_{1})^{2}}  & \frac{(u-v)^2}{ (u-v_{1})^{2}} \cS -\frac{2 (u-u_{-1}) (v-v_{1})}{(u-v_{1}) (u_{-1}-v)}
 \end{array} } \right)-2 K^{(-1)} (\cS-1)^{-1} Q^{(-1)},
\end{eqnarray*}
where 
\[
K^{(-1)}= \left(\begin{array}{c} \frac{(u-v)(u_{-1}-u)}{u_{-1}-v}  \\
 \frac{(u-v) (v-v_1)}{u-v_1}  \end{array}\right),\quad 
Q^{(-1)}=\left(\begin{array}{cc} \frac{v-v_{1}}{(u-v) (u-v_{1})}, & \frac{u-u_{-1}}{(u-v) (u_{-1}-v)}  \end{array}\right).
\]
Thus, the Heisenberg ferromagnet lattice has infinitely many local symmetries $\cR^l (K^{(1)})$ and  $\cR^{-l} (K^{(-1)})$ for all $l\in \bbbn$.
The phenomenon has been explored for the Ablowitz-Ladik lattice and the Bruschi-Ragnisco lattice in \cite{mr93c:58096}. In this paper,
we compute the weakly nonlocal inverse recursion operators for all multi-component integrable differential-difference equations if existing.
Such inverses do not exist for scalar nonlinear integrable differential-difference equations. 
For a given weakly nonlocal difference operator, to answer whether there exists a weakly nonlocal inverse operator is still an open problem.

The arrangement of this paper is as follows: First we review two closely related topics concerning the Lax representations. One is the Darboux transformations of 
the Lax representation, from which we derive the integrable differential-difference equations. Another topic is to derive
the recursion operator for the resulting equations using the Darboux transformation. We illustrate the methods through
two typical examples: the well-known nonlinear Schr\"odinger equation and a deformation of the derivative nonlinear
Schr\"odinger equation corresponding to the dihedral reduction group $\bbbd_2$.

We complete the paper with a long list
of integrable differential-difference equations, where we list equations themselves, their Hamiltonian structures, recursion operators, nontrivial
generalised symmetries and their Lax representations. We also include partial results on their master symmetries.
For some equations we add further notes concerning the links with other known equations and the weakly nonlocal inverses of recursion 
operators if existing. The list is far from being complete.

We mainly refer to the
source where we learned about each system although some attempt has been made to track the
original contributions. 
Compiling the list we have verified and made consistent the objects collected from the vast literature. Our list also includes a number of new results (to the best of our knowledge):
\begin{itemize}
\item The Hamiltonian operators, symplectic operators and recursion operators given in section \ref{yamivol} for equations (\ref{v1})--(\ref{v3}), and the relations among them;
\item The Hamiltonian operators, symplectic operators and recursion operators for 
the Kaup-Newell lattice (section \ref{seckn}), the Chen-Lee-Liu lattice (section \ref{seccll}) and the Ablowitz-Ramani-Segur (Gerdjikov-Ivanov) lattice
(section \ref{secgi});
\item All weakly nonlocal inverse recursion operators if existing (except for the cases of Ablowitz-Ladik lattice and Bruschi-Ragnisco lattice, which were known).
\end{itemize}

\section{Lax representations and Darboux matrices}\label{Sec2}
With an evolutionary nonlinear partial differential system 
\begin{equation}\label{pde}
 \bu_{t}=\bF(\bu,\bu_x,\ldots \bu_{x\cdots x}), \qquad \bu\in\bbbc^m
\end{equation}
solvable via the spectral transform method \cite{zmnp,AS,mr93g:35108} one associates a pair of linear operators
\[
 L=D_x-U(\bu;\lambda),\qquad A=D_{t}-V(\bu;\lambda),
\]
which is conventionally called the {\em Lax pair}.
Here $U, V$ are square matrices, whose entries are functions of the dependent variable $\bu $ and its $x$-derivatives and certain rational (in some cases elliptic) functions of the spectral parameter $\lambda$, such that 
equation (\ref{pde}) is equivalent to the condition of commutativity of these operators
\begin{equation}\label{zeroc}
  [L,\ A]=D_{t}(U)-D_x(V)+[U,V]=0.
\end{equation}
The latter is often called a {\em zero curvature representation} or {\em Lax representation} of equation (\ref{pde}).  
In this paper we mainly consider $U$ and $V$ to be $2\times 2$ matrices and their 
entries to be polynomial in spectral parameter $\lambda$.

Generally speaking, symmetries of an evolutionary equation are its compatible evolutionary equations.
Integrable equation (\ref{pde}) has an infinite sequence of commuting symmetries
\begin{equation}\label{sym}
 \bu_{t_k}=\bF^k(\bu,\bu_x,\ldots \bu_{x\cdots x}), \qquad k\in\bbbn\, ,
\end{equation}
which can be associated with a commutative algebra of linear operators 
\begin{equation}\label{LAAk}
 A^k=D_{t_k}-V^k(\bu;\lambda),\qquad [A^i,\ A^j]=0 . 
\end{equation}
Similar to equation (\ref{pde}) system (\ref{sym}) is equivalent to $[L,\ A^k]=0$.
Operator $A$ and equation (\ref{pde}) can be considered as members in the sequence of operators $\{A^k\}$ 
and symmetries (\ref{sym}) respectively, for  particular values of $k$.
The commutativity of operators  can be seen as a compatibility 
 condition for the infinite sequence of linear problems
\begin{equation}\label{lax0}
 D_x(\Psi)=U(\bu; \lambda)\Psi,\qquad D_{t_k}(\Psi)=V^k(\bu; \lambda)\Psi,
\end{equation}
i.e. the condition for the existence of a common fundamental solution  $\Psi$ of all these problems, $\det \Psi\ne 0$.

To Darboux transformations we refer as a linear map $\cS$ acting on a fundamental solution 
\begin{equation}\label{SPsi}
\cS:\ \Psi\mapsto \overline{\Psi}=M\Psi,\qquad \det\,M\ne 0
\end{equation}
such that the matrix function $\overline{\Psi}$ is a fundamental solution of the linear problems 
\begin{equation}\label{lax1}
 D_x(\overline{\Psi})=U(\overline{\bu}; \lambda)\overline{\Psi},\qquad D_{t_k}(\overline{\Psi})=V^k(\overline{\bu}; \lambda)\overline{\Psi},
\end{equation}
with new ``potentials''  $\overline{\bu}$. 
The matrix $M$ is often called the Darboux matrix.
The entries of the  Darboux matrix $M$ are a rational (elliptic) functions of the spectral parameter 
$\lambda$. As a function of $\lambda$ the determinant of $M$ may vanish only at a finite set points on the Riemann sphere 
(the parallelogram of periods).
The Darboux matrix $M$ depends on $\bu,\bar \bu$ and may also depend on some auxiliary functions 
$\bg$  (or parameters, if $\bg$ is a constant). For examples of such $\bg$, we refer to the matrices in (\ref{MNK}).
So we denote the Darboux matrix by $M=  M(\bu,\overline{\bu},\bg ;\lambda)$.

From the compatibility of (\ref{SPsi}) and (\ref{lax1}) it follows that 
\begin{eqnarray}
&& D_x(M)=U(\overline{\bu}; \lambda)M-MU(\bu; \lambda),\label{backlundx}\\
&& D_{t_k}(M)=V^k(\overline{\bu}; \lambda)M-MV^k(\bu; \lambda).\label{backlundtk}
\end{eqnarray}
Equations (\ref{backlundx}) and (\ref{backlundtk}) are differential equations which relate two solutions  $\bu$ 
and $\bar\bu$ of  (\ref{pde}) and (\ref{sym}). In the literature they are also often called {\sl B\"acklund transformations}. 

A Darboux transformation maps one compatible system (\ref{lax0}) into another one (\ref{lax1}). It defines a 
Darboux map $\cS: \bu\mapsto\overline{\bu}$.
The map (\ref{SPsi}) is invertible ($\det\,M\ne 0$) and it can be iterated
\[
  \cdots\acS \Psi={M}(\underline{\bu},\bu,\underline{\bg};\lambda)\underline{\Psi}\acS\overline{\Psi}=
  M(\bu,\overline{\bu},\bg;\lambda)\Psi\acS\overline{\overline{\Psi}}=
  {M}(\overline{\bu},\overline{\overline{\bu}},\overline{\bg};\lambda)\overline{\Psi}
  \acS\overline{\overline{\overline{\Psi}}}=
 {M}(\overline{\overline{\bu}},\overline{\overline{\overline{\bu}}},\overline{\overline{\bg}};\lambda) \overline{\overline{\Psi}}\acS\cdots\, .
\]
It suggests notations
$$
 \ldots \Psi_{-1}=\underline{\Psi},\ \Psi_0=\Psi,\ \Psi_1=\overline{\Psi},\ \Psi_2=\overline{\overline{\Psi}}, \ldots,
$$
$$ \ldots \bu_{-1}=\underline{\bu},\ \bu_0=\bu,\ \bu_1=\overline{\bu},\ \bu_2=\overline{\overline{\bu}}, \ldots,$$
$$ \ldots \bg_{-1}=\underline{\bg},\ \bg_0=\bg,\ \bg_1=\overline{\bg},\ \bg_2=\overline{\overline{\bg}}, \ldots\, .$$
With a vertex $k$  of the one dimensional lattice $\bbbz$ we associate variables $\Psi_k$ and $ \bu_k$; 
with the edges joining vertices $k$ and $k+1$  we associate the auxiliary functions (parameters) $\bg_k$ and the matrix $M_k=M(\bu_k,\bu_{k+1},\bg_k;\lambda)$. In this notations
the Darboux maps $\cS$ and $\cS^{-1}$ increases and decreases the subscript index by one, and therefore we shall call it  as $\cS$--shift, or shift operator $\cS$.
In what follows we often shall omit zero in the subscript index and write  $\bu,\bg$ instead of $\bu_0,\bg_0$. 

In these notations (\ref{backlundx}) and the sequence (\ref{backlundtk}) are a hierarchy of compatible systems of differential difference equations. 
When the resulting equations from (\ref{backlundx}) and (\ref{backlundtk}) are in the evolutionary form they
form an infinite dimensional Lie algebra of commuting symmetries. The existence of an infinite algebra of 
commuting symmetries is often taken as a definition of integrability of the equation (and of the whole hierarchy of symmetries) \cite{mr93b:58070, mr89g:58092, asy, mr99g:35058}.

In order to illustrate this construction we consider two examples:  (1) the well known example of the 
nonlinear Schr\"odinger equation; (2) the new results on
differential difference equations corresponding to the dihedral reduction group $\bbbd_2\simeq \bbbz_2\times\bbbz_2$ (the group of Klein).
In the next section we will use these examples for illustration of the derivation of recursion operators.
 
\subsection{The nonlinear Schr\"odinger  equation}\label{sec21} 
The nonlinear Schr\"odinger (NLS) equation
\begin{equation}\label{nls}
\begin{array}{l}
 ×2p_t={\phantom{ - }}p_{xx}-8p^2q\\
 2q_t=-q_{xx}+8q^2p
\end{array}
\end{equation}
has a zero curvature representation (\ref{zeroc}) where \cite{mr53:9966}
\begin{eqnarray}
U(\bu;\lambda)&=&\left(\begin{array}{rr}
        0& 2p\\2q&0
       \end{array}\right)+\lambda\left(\begin{array}{rr}
        1& 0\\0&-1
       \end{array}\right),\qquad \quad \bu=\left(\begin{array}{l}
                                            p\\q
                                           \end{array}\right),\\ 
V(\bu;\lambda)&=&\left(\begin{array}{rr}
        -2pq& p_x\\-q_x&2pq
       \end{array}\right)+\lambda\left(\begin{array}{rr}
        0& 2p\\2q&0
       \end{array}\right)+\lambda^2\left(\begin{array}{rr}
        1& 0\\0&-1
       \end{array}\right).
\end{eqnarray}
The NLS equation has an infinite hierarchy of commuting symmetries. The matrix part of the corresponding linear operators $A^k=D_{t_k}-V^k$ 
have the form
\begin{equation}\label{V0-Vk}
V^0=J,\qquad  V^{k+1}=\lambda V^k+B^k(\bu)
\end{equation}
where $J=\mbox{diag}(1,-1)$  and $B^k(\bu)$ are  traceless matrices with entries depending on $p,q$ and their $x$-derivatives. 
Matrices $B^k(\bu)$ can be found recursively \cite{mr2001h:37146} and in particular we have
\begin{equation}
 \label{V1-3}
 V^1=U(\bu;\lambda),\qquad V^2=V(\bu;\lambda)\qquad V^3=\lambda V(\bu;\lambda)+\frac{1}{2}\left(\begin{array}{cc}
                      2 pq_x-2qp_x&p_{xx}-8p^2q\\q_{xx}-8q^2p&2qp_x-2pq_x
                                                                                     \end{array}\right).
\end{equation}

Symmetries corresponding to $A^0,A^1,A^2$ and $A^3$ are
\begin{equation}\label{nlssym}
 \begin{array}{l}
  ×p_{t_0}={\phantom{ - }}2p\\
  q_{t_0}=-2q
 \end{array}\qquad 
  \begin{array}{l}
  ×p_{t_1}=p_x\\
  q_{t_1}=q_x
\end{array}\qquad 
\begin{array}{l}
  ×p_{t_2}={\phantom{ - }}\frac{1}{2}p_{xx}-4p^2q\\
  q_{t_2}=-\frac{1}{2}q_{xx}+4q^2p
\end{array}\qquad 
\begin{array}{l}
  ×p_{t_3}=\frac{1}{4}p_{xxx}-6pqp_x\\
  q_{t_3}=\frac{1}{4}q_{xxx}-6pqq_x.
\end{array}
\end{equation}

It is known (see \cite{adler_dis}, \cite{poland09}) that any Darboux matrix for the NLS equation is 
a composition of the following three elementary Darboux matrices:
\begin{equation}\label{MNK}
 M(\bu,\bu_1,f;\lambda)=\left(\begin{array}{cc}
                               \lambda+f&p\\q_1&1
                              \end{array}\right),\
N(\bu,\bu_1,h;\lambda)=\left(\begin{array}{cc}
                               \lambda+h&p\\p^{-1}&0
                              \end{array}\right),\
K(\bu,\bu_1,\alpha;\lambda)=\left(\begin{array}{cc}
                              \alpha&0\\0&\alpha^{-1}
                              \end{array}\right)
\end{equation}
and their inverses. 

The Darboux map and/or the corresponding differential difference equation for each above elementary Darboux matrix can be derived as follows.

($K$): The  map with the Darboux matrix $K$ does not depend on the spectral parameter 
$\lambda$ and variables $\bu,\bu_1$. Moreover, it follows from (\ref{backlundx})   that $\alpha$ is a constant (does not depend on $x$) 
and $p_1=\alpha^{2}p,q_1=\alpha^{-2} q$. Thus in this case the Darboux map is a gauge transformation corresponding to 
a point symmetry of the equation (\ref{nls}) and its hierarchy 
of symmetries.

($N$): Substitution of $N(\bu,\bu_1,h;\lambda)$ in (\ref{backlundx}) yields
\begin{equation}\label{nlsd1}
 q_1=\frac{1}{p},\qquad h_x=2p_1q_1-2pq,\qquad p_x=-2hp,\qquad q_{1,x}=2hq_1
\end{equation}
and therefore we have got the following explicit  Darboux map 
\[
 p_1=p^2q-\frac{p}{4}\left(\frac{p_x}{p}\right)_x,\qquad q_1=\frac{1}{p}.
\]
After the change of the variables $p=\exp (\phi)$ (and thus $q=\exp (-\phi_{-1})$) equations (\ref{nlsd1}) yields the following system of evolutionary equations
\begin{equation}\label{todaev}
 \phi_x=-2h,\qquad h_x=2\exp(\phi_{1}-\phi)-2\exp(\phi-\phi_{-1})
\end{equation}
which after the elimination of variable $h$ takes form of the
Toda lattice
\[
 \phi_{xx}=4\exp(\phi-\phi_{-1})-4\exp(\phi_{1}-\phi).
\]
It is an infinite chain of differential equations for dependent variables $\phi_n,\ n\in \bbbz$.

Notice that $\phi=\ln p$ and $h=-\frac{\phi_x}{2}$. Using (\ref{nlsd1}) to eliminate $x$-derivatives from (\ref{nlssym}) one can obtain symmetries of the Toda chain (\ref{todaev}):
\begin{equation}\label{todasym}
 \begin{array}{l}
  \phi_{t_0}=2\\
  h_{t_0}=0
 \end{array}\qquad 
  \begin{array}{l}
  ×\phi_{t_1}=-2h\\
  h_{t_1}=2(\cS-1)\exp(\phi-\phi_{-1})
\end{array}\qquad 
\begin{array}{l}
  \phi_{t_2}=2h^2-2(\cS+1)\exp(\phi-\phi_{-1})\\
  h_{t_2}=-2 (\cS-1)( \exp(\phi-\phi_{-1})(h_{-1}+h))
\end{array}
\end{equation}
\[
 \begin{array}{l}
  \phi_{t_3}=-2h^3+2\exp(\phi-\phi_{-1})(2h+h_{-1})+2\exp(\phi_1-\phi)(2h+h_{1})\\
  h_{t_3}=2 (\cS-1)((h_{-1}^2+h_{-1}h+h^2)\exp(\phi-\phi_{-1})+\exp(2\phi-2\phi_{-1})+(\cS+1)\exp(\phi-\phi_{-2}))  
\end{array}\ .
\]
Symmetries (\ref{todasym}) have Darboux-Lax representation
\begin{equation}
 \label{todad}
 N_{t_k}-\cS(U^k)N+NU^{k}=0
\qquad
       N=\left(\begin{array}{cc}
                               \lambda+h&\exp(\phi)\\\exp(-\phi)&0
                              \end{array}\right),      
\end{equation}
and matrices $U^0,\ldots ,U^3$ are obtained from $V^0,\dots,V^3$ given by (\ref{V0-Vk}) and (\ref{V1-3}) by elimination of $x$-derivatives using (\ref{nlsd1}).
In the next section we shall derive a recursion operator for generating symmetries of the Toda lattice.

($M$): Substitution of $M(\bu,\bu_1,f;\lambda)$ in (\ref{backlundx}) leads to the following  system of differential-difference equations:
\begin{equation}\label{nlsd2}
f_x=2p_1q_1-2pq,\qquad p_x=2p_1-2fp,\qquad q_{x}=-2q_{-1}+2f_{-1}q.
\end{equation}
Symmetries of this system can be found from the Lax Darboux representations (\ref{backlundtk}) where $V^k$ given by (\ref{V0-Vk}) and (\ref{V1-3}) 
after elimination of $x$-derivatives with the help of (\ref{nlsd2}):
\begin{equation}\label{nlssymd2}
 \begin{array}{l}
  ×p_{t_0}={\phantom{ - }}2p\\
  q_{t_0}=-2q\\
  f_{t_0}=0
 \end{array}\qquad 
  \begin{array}{l}
  ×p_{t_1}=2p_1-2fp\\
  q_{t_1}=-2q_{-1}+2f_{-1}q\\
  f_{t_1}=2p_1q_1-2pq
\end{array}\qquad 
\begin{array}{l}
  ×p_{t_2}=2 (f^2 p - f p_1 - f_1 p_1  - 
   p^2 q - p p_1 q_1+ p_2)\\
  q_{1,t_2}=2 (  f_{-1} q +f q - 
   f^2 q_1 + pq q_1 + p_1q_1^2-q_{-1})\\
  f_{t_2}=2(\cS-1) (p  q_{-1}+p_1q - (f_{-1}+f) p q )
\end{array}
\end{equation}
System (\ref{nlsd2}) and its symmetries (\ref{nlssymd2}) have first integral $\Phi=f-pq_{-1}$, so that $\Phi_{t_k}=0$. Indeed, $\det M(\bu,\bu_1,f;\lambda)=\lambda+f-pq_1$ should be a constant 
(does not depend on $x, t_k$) because the matrices $U,V^k$ are traceless (Abel's Theorem). Thus 
one can set $f_k=p_k q_{k+1}+\alpha_k$, 
where $\alpha_k\in\bbbc$ is a constant. We can eliminate $f$ from the system (\ref{nlsd2}) and it leads to
\begin{equation}\label{nlsd2nof}
 p_x=2p_1-2p^2q_1-2\alpha p,\qquad q_{x}=-2q_{-1}+2p_{-1}q^2+2\alpha_{-1}q,
\end{equation}
whose symmetries can be obtained from (\ref{nlssymd2}) by the same elimination of $f$.
When $\alpha_k=0$, equation (\ref{nlsd2nof}) becomes the Merola-Ragnisco-Tu lattice under an invertible transformation 
$x=\frac{1}{2} t$, $p=u$ and $q_1=v$ listed in Section \ref{merola}.

\subsection{Equations corresponding to the dihedral reduction group $\bbbd_2\simeq \bbbz_2\times\bbbz_2$}

Integrability of the  equation
\begin{equation}\label{d2nls}
\begin{array}{l}
 ×2p_t={\phantom{ - }}p_{xx}+4q_x-8(p^2q)_x\\
 2q_t=-q_{xx}+4p_x-8(q^2p)_x
\end{array}
\end{equation}
has been established in \cite{ mr89g:58092}. This equation can be seen as a non-trivial inhomogeneous 
deformation of the {\sl derivative nonlinear Schr\"odinger} equation. The corresponding Lax pair
\[ L(\bu;\lambda)=D_x-V^1(\bu;\lambda),\qquad A(\bu;\lambda)=D_t-V^2(\bu;\lambda)\]
has the matrix part of the form
\begin{eqnarray}
&&V^1(\bu;\lambda)=2p\ba_1(\lambda)+2q\ba_2(\lambda)+2\ba_3(\lambda)\label{d2L1}\\
&&V^2(\bu;\lambda)= w(\lambda) V^1(\bu;\lambda)+\frac{p_x-4p^2q}{2}\ba_1(\lambda)-\frac{q_x+4q^2p}{2}\ba_2(\lambda)  -2pq \ba_3(\lambda)\label{d2L2}
\end{eqnarray}
where
\begin{equation}\label{a123}
\ba_1(\lambda)= \left(\begin{array}{cc}
        0& \lambda\\ \lambda^{-1}&0
       \end{array}\right),\ 
\ba_2(\lambda)= \left(\begin{array}{cc}
        0& \lambda^{-1}\\ \lambda&0
       \end{array}\right),\ 
\ba_3(\lambda)=\frac{\lambda^2-\lambda^{-2}}{2}\left(\begin{array}{rr}
        1& 0\\0&-1
       \end{array}\right) .
\end{equation}

Matrices $V^k(\bu;\lambda)$ are invariant with respect to the group of transformation (the reduction group \cite{mik81}) 
\[
 \left(\begin{array}{rr}
        1& 0\\0&-1
       \end{array}\right)V^k(\bu;-\lambda)\left(\begin{array}{rr}
        1& 0\\0&-1
       \end{array}\right)=V^k(\bu;\lambda),\quad \left(\begin{array}{rr}
        0& 1\\1&0
       \end{array}\right)V^k(\bu;\lambda^{-1})\left(\begin{array}{rr}
          0& 1\\1&0
       \end{array}\right)=V^k(\bu;\lambda),
\]
which is isomorphic to the  dihedral group $\bbbd_2\simeq \bbbz_2\times\bbbz_2$ (the group of Klein). Function 
\begin{equation}\label{wd}
  w(\lambda)=\frac{\lambda^2+\lambda^{-2}}{2}\,
\end{equation}
 is a primitive automorphic function of the corresponding M\"obius group ($ w(\lambda)=w(-\lambda)=w(\lambda^{-1})$)
and any rational automorphic function of this group is a rational function of $w(\lambda)$ \cite{LM05}. A hierarchy of higher symmetries of 
equation (\ref{d2nls}) can be generated by the Lax operators $A^k=D_{t_k}-V^k$, where the matrices $V^k$ are of the form
\begin{equation}\label{vk1d}
 V^{k+1}=w(\lambda)V^k+r_1^k \ba_1(\lambda)+r_2^k \ba_2(\lambda)+r_3^k \ba_3(\lambda)
\end{equation}
and the coefficients $r_1^k,r_2^k,r_3^k$ are  polynomials in $p,q$ and their $x$-derivatives and they can be found recursively.

It has been shown in \cite{smx} that an elementary Darboux matrix for the Lax operator (\ref{d2L1}) can be written in the form
\begin{equation}
\label{Md}
 M(p,q_1,f,g;\lambda)=f (w(\lambda) \bI+\ba_3(\lambda)+p\ba_1(\lambda)+q_1\ba_2(\lambda)+g \bI)
\end{equation}
where $\bI$ is a unit matrix. Matrices $V^k$ are all traceless and therefore the determinant 
\[\det M(p,q_1,f,g;\lambda)=f^2 \left(2 w(\lambda)  ( g-  p q_1)+ 1 +  g^2 -  p^2 -  q_1^2\right) =2  w(\lambda) \Phi^1+\Phi^2
\]
does not depend on $x$ and $t_k$. Thus we have two invariants 
\begin{equation}\label{fi12} 
 \Phi^1=f^2 ( g-  p q_1),\qquad \Phi^2=f^2(1 +  g^2 -  p^2 -  q_1^2)\, .
\end{equation}
By choosing an appropriate scaling $M\to \gamma M$ we can make  $\Phi^1=1$ (if $\Phi^1\ne 0$) or $\Phi^2=1$ (if $\Phi^2\ne 0$).

There are three essentially different cases \cite{smx}:
\begin{enumerate}
 \item[I.] $\Phi^1=0,\ \Phi^2=1$, in this case the determinant is a constant (does not depend on $\lambda$);
 \item[II.] $\Phi^1=1,\ \Phi^2=\pm 2 $, in this case the determinant has two double zeroes and is a square of a rational function of $\lambda$; 
 \item[III.] $\Phi^1=1,\ \Phi^2=2 \alpha$, the determinant has four distinct zeros in the complex plane $\lambda$ (assuming that $\alpha\ne\pm 1$).
\end{enumerate}

In the case I we have 
\begin{equation}\label{d2case0}
  g=pq_1
\end{equation}
and $f$ can be found as a solution of equation $f^2(1 -  p^2)(1 -  q_1^2)=1$. Thus we can determine the matrix  $M(p,q_1,f,pq_1;\lambda)$ in (\ref{Md}).
Substituting it and $V^1(\bu;\lambda)$ given by (\ref{d2L1}) instead of $U(\bu; \lambda)$ into  (\ref{backlundx}) leads to
the system
\begin{equation}\label{d2mv}
p_x=2(1-p^2)(q_1-q),\qquad q_x=2(1-q^2)(p-p_{-1}),
\end{equation}
which can be written as a scalar equation for one function $v_{2n}=p_n,\ v_{2n-1}=q_n$
\begin{equation}\label{mvoleq}
 v_x=(1-v^2)(v_1-v_{-1}).
\end{equation}

In the case II we shall take $2\Phi^1+\Phi^2=0$ (the other choice of the sign would eventually lead to a point equivalent system). We have
\[
 2\Phi^1+\Phi^2=f^2 (1 + g + p + q_1)(1 + g - p - q_1) =0
\]
Let us choose $1 + g + p + q_1=0$ (the second choice $1 + g - p - q_1=0$ would lead to a point equivalent system). Then the substitution of 
$M(p,q_1,f,-p-q_1-1;\lambda)$ and $V^1(\bu;\lambda)$ given by (\ref{d2L1}) instead of $U(\bu; \lambda)$ into    (\ref{backlundx}) leads to
the system
\begin{equation}\label{d2c2a}
p_x=2(1+p)(\cS-1)(q-p-pq),\qquad q_{1,x}=2(1+q_1)(\cS-1)(p-q-pq).
\end{equation}

In the generic case III of the elementary Darboux transformation we have
\begin{equation}\label{d2fg}\def \bF {{\bf F}}
\Phi^2-2\alpha \Phi^1=(  g^2-2 \alpha g -  p^2 -  q_1^2+2 \alpha p q_1+1)f^2=0,\qquad f=\frac{1}{\sqrt{ g-  p q_1}} .
\end{equation}
The substitution of 
$M(p,q_1,f,g;\lambda)$ and $V^1(\bu;\lambda)$ given by (\ref{d2L1}) instead of $U(\bu; \lambda)$ into  (\ref{backlundx}) leads to
the system
\begin{eqnarray*}
 p_x&=&2g (p_1-p)+2q_1-2q-2p(p_1q_1-pq),\\
 q_{1,x}&=&2g (q_1-q)+2p_1-2p-2q_1(p_1q_1-pq),\\
 g_x&=&2p (p_1-p)+2q_1(q_1-q)-2g(p_1q_1-pq),\\
 f_x&=&2f(p_1q_1-pq).
\end{eqnarray*}
The invariants $\Phi^1,\Phi^2$ are the first integrals of this system. Thus the functions $f$ and $g$ can be eliminated using 
the first integrals (\ref{d2fg}). Note that in (\ref{d2fg}) the parameter $\alpha$ is a constant in $x,t_k$, but it may depend on the shift variable, so that 
$\cS(\alpha)=\alpha_1$.

\section{Recursion operators for differential-difference equations}

In this section we show how to derive a recursion operator using a Darboux-Lax representation for a  differential-difference equation.
Our construction is a natural generalisation of the method used in the theory of integrable PDEs \cite{ mr2001h:37146, zhang02, DS08, wang09, GGMT}.
The main idea of the method is based on the fact that matrices $V^k(\bu,\lambda)$ of the operators $A^k=D_{t_k}-V^k(\bu,\lambda)$ corresponding to a hierarchy 
can be related as
\begin{equation}\label{VK}
 V^{k+1}(\bu,\lambda)=\mu(\lambda)V^k(\bu,\lambda)+B^k(\bu,\lambda),
\end{equation}
where $\mu(\lambda)$ is a rational (elliptic in the case of the Landau-Lifshitz equation) multiplier and $B^k(\bu,\lambda)$ is a rational matrix with a fixed (i.e. $k$ independent) divisor 
of poles. If the system and its Lax representation is obtained as a result of a reduction with a reduction group $G$, then the multiplier  $\mu(\lambda)$ is a primitive automorphic function 
\cite{LM05} of a finite reduction group, or  in the elliptic case is one of the  generators of the $G$-invariant 
subring of the coordinate ring \cite{DS08}. The matrix  $B^k(\bu,\lambda)$ also depends on the dependent variables $\bu$ and its $x$-derivatives.

The substitution of (\ref{VK}) in the Lax representation $[L,A^k]=0$  (\ref{LAAk}) results in 
\begin{equation}\label{BK}
D_{t_{k+1}}(L)=\mu(\lambda) D_{t_{k}}(L)-B^k L+L B^k.
\end{equation}
One can use equation (\ref{BK}) to find $B^k$ in terms of variables $\bu, \bu_{t_k}, \bu_{t_{k+1}}$ and  $x$ derivatives of $\bu$. Then equation    (\ref{BK})
can be regarded as a recursive relation $\bu_{t_{k+1}}=\cR (\bu_{t_{k}})$, where $\cR$ is a linear pseudo-differential recursion operator mapping a symmetry
to a new symmetry.
A recursion operator can be related to bi-Hamiltonian structure. Indeed, if $\cH_1$ and $\cH_2$ are two compatible Hamiltonian operators,
then $\cR=\cH_1\cH_2^{-1}$ is a Nijenhuis recursion operator \cite{mr82g:58039, mwx2}. 
The sufficient condition for $\cR$ to be a recursion operator is \cite{mr58:25341}
\[
D_{t_k}(\cR)=[\bF^k_{*},\cR],
\]
where $\bF^k_{*}$ is the Fr\'echet derivatives of $ \bF^k$.

A similar construction can be used in the  differential difference case \cite{wang12}. The substitution of (\ref{VK}) in the Darboux-Lax representation   (\ref{backlundtk}) results in 
\begin{equation}\label{MBK}
D_{t_{k+1}}(M)=\mu(\lambda) D_{t_{k}}(M)-\cS(B^k) M+M B^k.
\end{equation}
Equation (\ref{MBK}) enables us to express the entries of the matrix $B^k$ in terms of  $\bu, \bu_{t_k}, \bu_{t_{k+1}}$ and their $\cS$-shifts. 
It enables us to find a linear pseudo-difference operator $\cR$ such that $\bu_{t_{k+1}}=\cR (\bu_{t_{k}})$, i.e. a recursion operator for a differential-difference
hierarchy of commuting symmetries. Like the differential case, if we know two compatible Hamiltonian operators $\cH_1, \cH_2$ or a compatible pair of a Hamiltonian 
 $\cH$ and a symplectic $\J$ operators then $\cR=\cH_1\cH_2^{-1}$ and $\tilde \cR=\cH \J$ are recursion operators. From this construction it follows that a Darboux 
 matrix $M$ and a multiplier $\mu(\lambda)$ defines a recursion operator completely and uniquely.

In this section we shall illustrate this construction on a few examples. In the next section we shall present an extensive (but far not complete) 
list of integrable differential-difference equations with recursion operators, multi-Hamiltonian structure and Darboux-Lax representations.

\subsection{Differential difference equations from the NLS equation}
 We are going to illustrate the construction  using the Darboux matrices $M$ and $N$ (\ref{MNK}) 
related to the NLS equation. In this case the multiplier $\mu(\lambda)=\lambda$ (\ref{V0-Vk}) and the matrix $B^k$ does not depend on the spectral parameter $\lambda$.

(N): Let us construct a recursion operator for the Toda lattice (\ref{todaev}) using  the Darboux matrix $N$  (\ref{todad}) and the multiplier $\mu(\lambda)=\lambda$. We substitute
\begin{equation}
 \label{BB}
 B=\left(\begin{array}{rr}
   a&b\\c&-a
         \end{array}
\right)
\end{equation}
in the equation 
\begin{equation}
 \label{trs}
D_{t_{k+1}}(N)=\lambda D_{t_{k}}(N)-\cS(B) N+N B.
\end{equation}
Linear part in $\lambda$ of (\ref{trs}) leads to the system of equations
\[
 (\cS-1)a= h_{t_k}, \quad b= -\phi_{t_k}\exp(\phi),\quad c_1= -\phi_{t_k}\exp(-\phi).
\]
Thus we can find all coefficients  $a= (\cS-1)^{-1}h_{t_k},\ b= -\exp(\phi)\phi_{t_k},\ c= -\exp(-\phi_{-1})\cS^{-1}\phi_{t_k}$ of the matrix $B$.
Then the $\lambda$ independent part of  (\ref{trs}) reduces to equations
\begin{eqnarray*}
&&\phi_{t_{k+1}} = b h-a-a_1=-(\cS+1)(\cS-1)^{-1}h_{t_k} - h\phi_{t_{k}}, \\
&&h_{ t_{k+1}}= (a-a_1)h+c \exp(\phi)-b_1 \exp(-\phi)=-h h_{t_k} - 
  \exp(\phi-\phi_{-1}) \cS^{-1}\phi_{t_{k}} +  \exp(\phi_1-\phi) \cS\phi_{t_{k}},
\end{eqnarray*}
which leads to a recursion 
\[
 \left(\begin{array}{r}
\phi_{t_{k+1}}\\ h_{t_{k+1}}        ×
       \end{array}\right)=\cR\left(\begin{array}{r}
\phi_{t_{k}}\\ h_{t_{k}}        ×
       \end{array}\right)
\]
with the pseudo difference operator  
\begin{equation}\label{todarec}
\cR=\left(\begin{array}{cc}
           ×-h&-(\cS+1)(\cS-1)^{-1}\\ \exp(\phi_1-\phi) \cS-\exp(\phi-\phi_{-1}) \cS^{-1}&-h
          \end{array}\right).
\end{equation}
Starting from the seed symmetry $\phi_{t_0}=-2,\ h_{t_0}=0$ one can recursively produce the hierarchy of
symmetries (\ref{todasym}). 

Similar to the case of scalar discrete equations \cite{mwx2, mwx1},  the canonical series of the densities of local conservation laws can be found 
taking residues $\rho_k=\mbox{res}\ \cR^k$.  
In the case of matrix pseudo--difference operator $\cA$ a residue $  \mbox{res} \cA$ is defined as  follows:

Any pseudo--difference operator $\cA$ can be uniquely represented by its Laurent series $$\cA=\sum_{k=0}^\infty \cA^{m-k}\cS^{m-k},$$ then the residue is defined as 
$\mbox{res}\cA=\mbox{trace}(\cA^0)$. For example,  we rewrite the recursion operator (\ref{todarec}) as 
\begin{eqnarray*}
 \cR=\left(\begin{array}{cc}0&0\\ \exp(\phi_1-\phi) &0
          \end{array}\right)\cS+\left(\begin{array}{cc}
           -h&-1\\ 0&-h\end{array}\right)+\left(\begin{array}{cc}0 &-2\\ -\exp(\phi-\phi_{-1})&0
          \end{array}\right) \cS^{-1}+\cdots
\end{eqnarray*}
It follows that
\[\rho_1=\mbox{res}\cR=-2h,\ \rho_2=\mbox{res}\cR^2=2h^2-2\exp(\phi_1-\phi)-2\exp(\phi-\phi_{-1}), \ldots \ ,\]
are conserved densities for the Toda lattice (\ref{todaev}). Indeed, we have
\begin{eqnarray*}
&&D_x \rho_1=-4 (\cS-1) \exp(\phi-\phi_{-1});\\
&&D_x  \rho_2=4 (\cS-1) \left(\exp(\phi-\phi_{-1}) (h_{-1}+h)\right).
\end{eqnarray*}

(M): Now we take the Darboux matrix $M$ (\ref{MNK}) and the matrix $B$ of the form (\ref{BB}). From the linear part in $\lambda$  of the equation 
\begin{equation}
 \label{trs1}
D_{t_{k+1}}(M)=\lambda D_{t_{k}}(M)-\cS(B) M+M B
\end{equation}
it follows that 
\[a= (\cS-1)^{-1}f_{t_k},\quad b= -p_{t_k},\quad c= q_{t_k}.\]
The $\lambda$ independent part of (\ref{trs1}) leads to the recursion relation for the system (\ref{nlsd2})
\begin{equation}
\left(\begin{array}{l}
       × f_{t_{k+1}}\\ p_{t_{k+1}}\\ q_{t_{k+1}}
      \end{array}\right)=\left(\begin{array}{l}

-f f_{t_k} +\cS(q p_{t_k} ) + p q_{t_k}, \\
-2 p(\cS-1)^{-1}f_{t_k} - f_{t_k} p - f p_{t_k} + \cS(p_{t_k}), \\
  2  q(\cS-1)^{-1}f_{t_k} - q\cS^{-1}(f_{t_k})  + \cS^{-1}q_{t_k} - f_{-1} q_{t_k} \end{array}\right)=\cR\left(\begin{array}{l} × f_{t_{k}}\\ p_{t_{k}}\\ q_{t_{k}}
      \end{array}\right)
\end{equation}
where a pseudo-difference recursion operator is of the form
\[
 \cR=\left(\begin{array}{ccc}
            -f&q_1\cS&p\\-p&-f+\cS&0\\-q\cS^{-1}&0&-f_{-1}+\cS^{-1}
           \end{array}\right)+
\left(\begin{array}{c}
   0\\-2p\\2q         
           \end{array}\right)(\cS-1)^{-1}(1\ 0\ 0).
           \]

Using the first integral of the system (\ref{nlsd2}) and eliminating $f_k=p_k q_{k+1}+\alpha_k$ we obtain the corresponding recursion relation and operator for the system (\ref{nlsd2nof}) as follows:
\begin{equation}
\left(\begin{array}{l}
       p_{t_{k+1}}\\ q_{t_{k+1}}
      \end{array}\right)=\cR'\left(\begin{array}{l}  p_{t_{k}}\\ q_{t_{k}}
      \end{array}\right),
\end{equation}
where
 \[
 \cR'=\left(\begin{array}{cc}
            \cS-2pq_1-\alpha&-p^2 \cS -2 p p_{-1}\\-q^2\cS^{-1}&\cS^{-1}-\alpha_{-1}
           \end{array}\right)+
\left(\begin{array}{c}
   -2p\\2q         
           \end{array}\right)(\cS-1)^{-1} (q_1,\ p_{-1}).
\]
As we point out that equation (\ref{nlsd2nof}) is related to the Merola-Ragnisco-Tu lattice,
we can check that we obtain the same operator from the recursion operator for the Merola-Ragnisco-Tu lattice in Section \ref{merola} 
using formula (\ref{trar}).

\subsection{Difference equation corresponding to the dihedral reduction group}\label{sec32}
In this section, we demonstrate how to compute a recursion operator for equation (\ref{d2c2a}), whose Lax representation is invariant under the dihedral reduction group.

After a simple change of variables $p\to p-1,q\to q-1$ equation (\ref{d2c2a}) becomes
\begin{equation}
 \label{2dd}
p_x=2p(\cS-1)(2q-pq),\qquad q_{1,x}=2q_1(\cS-1)(2p-pq).
 \end{equation}
This equation is the relativistic Volterra lattice in Section \ref{rvol} under the scaling transformation  
\begin{eqnarray}\label{trar2}
p=-2 v,\quad  q=-2 u,\quad x=-\frac{t}{4} \ .
\end{eqnarray}
The  corresponding Darboux matrix  and the multiplier are
\begin{equation}\label{dar2d}
M=f\left(\begin{array}{cc}
\lambda^2+1-p-q_1&\lambda (p-1)+ \lambda^{-1}(q_1-1)     \\ 
\lambda (q_1-1) +\lambda^{-1}(p-1)   &  \lambda^{-2}+1-p-q_1
         \end{array}\right), \qquad \mu(\lambda)=\frac{1}{2}(\lambda^2+\lambda^{-2}).
\end{equation}
This information is sufficient to find a recursion operator, equation (\ref{2dd}) itself and its hierarchy
of local symmetries. It follows from
\[
 \mbox{det}M=-(\lambda-\lambda^{-1})^2 f^2 p q_1
\]
that $ f^2 p q_1$ does not depend on $x$ (Abel's Theorem). Thus we can set 
$
f=(pq_1)^{-\frac{1}{2}} \, 
$.

Expression  (\ref{vk1d}) suggests that 
\[
 B=a \ba_1+b \ba_2+c\ba_3\, .
\]
Now in equation (\ref{MBK}) the right and left hand sides are rational matrix functions in $\lambda$. The left hand side has only simple poles. Vanishing the coefficients at 
the third order poles in $\lambda^{-3}$ in the right hand side is equivalent to
\begin{equation}\label{abd1}
 \begin{array}{ll}
 a = \frac{1}{2} (c  + c_1)( p-1)+\frac{(p-1)p\cS q_{t_k}-(p+1)q_1 p_{t_k}}{4pq_1},\\
 b_1 = \frac{1}{2}( c+c_1)(q_1-1) +\frac{(q_1+1)p\cS q_{t_k}+(1-q_1)q_1 p_{t_k}}{4pq_1}.
 \end{array}
\end{equation}

The second order poles vanish iff
\begin{equation}
 \label{cd}
 c-c_1=\frac{q_1p_{t_k}+pq_{1,t_k}}{2pq_1}\, .
\end{equation}
Now we can simplify (\ref{abd1}) by eliminating $c_1$ and $c_{-1}$ from $a,b$
\begin{equation}\label{abd}
 \begin{array}{ll}
 a= c( p-1) -\frac{1}{2} p_{t_k},\\ 
 b= c( q-1) +\frac{1}{2} q_{t_k}.
 \end{array}
\end{equation}
The first order poles in equation (\ref{MBK}) vanishes if
\begin{equation}\label{recd}
 \begin{array}{ll}
  × p_{t_{k+1}}=&cp(\cS-1)(pq-2q)+(p-q_1-1-\frac{1}{2}p_1q_1)p_{t_k}-\frac{1}{2}pq_1\cS p_{t_k} +\\&(p-\frac{1}{2}p^2)q_{t_k}+(p-\frac{1}{2}pp_1)\cS q_{t_k},\\ \\
  q_{t_{k+1}}=&cq(\cS-1)(p_{-1}q_{-1}-2p_{-1})+(q+p_{-1}-1-\frac{1}{2}p_{-1}q_{-1})q_{t_k}-\frac{1}{2}p_{-1}q\cS^{-1} q_{t_k} +\\&(q-\frac{1}{2}q^2)p_{t_k}+
  (q-\frac{1}{2}qq_{-1})\cS^{-1} p_{t_k}.
 \end{array}
\end{equation}
The $\lambda $ independent part of equation (\ref{MBK}) is satisfied after the substitution of (\ref{cd}), (\ref{abd}) and (\ref{recd}) in  (\ref{MBK}).

Equation (\ref{recd}) together with (\ref{cd}) is a recurrence relation. The differential-difference equation (a seed symmetry) $p_x=2K^{(1)}_p,\ q_x=2K^{(1)}_q$ (\ref{2dd}) 
can be recovered from this recursion by taking the derivative 
of the right-hand side of equations (\ref{recd}) with respect to $c$, that is,
\[ (K^{(1)}_p,\ K^{(1)}_q)=(p(\cS-1)(pq-2q),\ q(\cS-1)(p_{-1}q_{-1}-2p_{-1})) .\]
It is not surprising. Indeed, in order to solve  equation (\ref{cd}) to find $c$ we need to invert the difference operator $\cS-1$, whose kernel is a 
field of constants. As a result, the vector $p_x,\ q_x$ may contribute to $p_{t_{k+1}},q_{t_{k+1}}$ with an
arbitrary constant coefficient. 

We now write explicitly down the recursion operator corresponding to the recursion (\ref{recd}) as follows:
\begin{eqnarray}
&&\cR=\left(\begin{array}{cc}
 p+q_1-1-\frac{1}{2}p_1q_1-\frac{1}{2}pq_1\cS&p-\frac{1}{2}p^2+(p-\frac{1}{2}pp_1)\cS\\
 q-\frac{1}{2}q^2+ (q-\frac{1}{2}qq_{-1})\cS^{-1} &q+p_{-1}-1-\frac{1}{2}p_{-1}q_{-1}-\frac{1}{2}p_{-1}q\cS^{-1}
           \end{array}\right) \nonumber\\
&&\qquad -\left(\begin{array}{c}p(\cS-1)(pq-2q)\\ q(\cS-1)(p_{-1}q_{-1}-2p_{-1})\end{array}\right)
           (\cS-1)^{-1}(\frac{1}{2p}\  \frac{1}{2q_1}\cS) \ .\label{repq}
\end{eqnarray}
Obviously $\cR+I$, where $I$ is the $2\times 2$ identity matrix, is also a recursion operator, which can be recovered from 
the recursion operator for the relativistic Volterra lattice listed in Section \ref{rvol} using the transformation (\ref{trar2}).

\section{A List of Integrable Differential-Difference Equations}\label{sec4}
In this section, we present a long list of integrable differential difference equations 
with their Hamiltonian structures, recursion operators, nontrivial
generalised symmetries and their Lax representations. 
To be self-contained, we introduce notations and remind some definitions of the objects in our list in 
terms of the Lie derivatives. The theoretic background and detail definitions of the Hamiltonian and symplectic operators 
can be found in monographs \cite{kp85, mr94j:58081}.
Here we would like to mention that there are some recent developments in the theory of nonlocal Hamiltonian structures for nonlinear partial differential equations in \cite{kac2013}.

Let $\bu=(u^1,\ldots, u^N)$ be a vector-valued function of variable $n\in \bbbz$ and time variable $t$. 
An evolutionary differential-difference equation of dependent variable $\bu$ is of the form
\begin{equation}\label{ddeq}
\bu_t=\bK[\bu], 
\end{equation} where $\bK [\bu]$ means that the smooth vector-valued function
$\bK$ depends on $\bu$ and its shifts $\bu_i=\cS^i \bu$, $i\in \mathbb{Z}$. 
In all our examples after an appropriate point change of variables
we can consider $\bK [\bu]\in \cF^N$ where  $\cF=(\bbbc,\bu,\cS)$ is a difference field of
rational functions of $\{u_k^i\, |\, k\in\bbbz, i=1,\ldots , N\}$.
A Fr\'echet derivative $a_\star$ of $a\in\cF$ is defined as a row vector of the difference operators 
\[
 a_\star=\sum_{k\in\bbbz}(\frac{\partial a}{\partial u_k^1},\ldots ,\frac{\partial a}{\partial u_k^N})\cS^k,
\]
thus the Fr\'echet derivative of $\bK [\bu]$ is a difference operator with square matrix coefficients 
and the entries of the coefficient matrices are elements of $\cF$.

A variational derivative of $a\in\cF$ is a column vector 
\[
 \delta_\bu(a)\coloneqq a_\star^\dagger (1)=\left(\frac{\partial}{\partial u^1},
\ldots ,\frac{\partial}{\partial u^N}\right) \sum_{k\in\bbbz}\cS^k(a).
\]

If equation (\ref{ddeq}) is Hamiltonian, then it can be written in the form
\[
 \bu_t=\cH (\delta_\bu (f))  
\]
where $\cH$ denotes a Hamiltonian (pseudo-)difference operator and $f$ is a Hamiltonian function (or the Hamiltonian of the system).


\begin{Def}\label{def}
Given a differential--difference evolutionary equation (\ref{ddeq}), we define that
\begin{itemize}
\item $\bG$ is its symmetry  if $L_{\bK} \! \bG\coloneqq [\bK, \bG]\coloneqq\bG_\star(\bK)-\bK_\star(\bG)=0$;
\item $\cH$ is its Hamiltonian operator if a Hamiltonian operator $\cH$ satisfies 
$$L_{\bK}\! \cH\coloneqq\cH_{\star}[{\bK}] -{\bK}_{\star} \cH-\cH {\bK}_{\star}^{\dagger}=0;$$
\item $\J$ is its symplectic operator if a symplectic operator $\J$ satisfies $$L_{\bK} \!\J\coloneqq\J_{\star}[{\bK}] +{\bK}_{\star}^{\dagger} \J+\J {\bK}_{\star}=0;$$
\item $\cR$ is its recursion operator if $L_{\bK}\! \cR\coloneqq{\cR}_{\star}[{\bK}] -{\bK}_{\star} \cR+\cR {\bK}_{\star}=0.$
\end{itemize}
Here $L_{\bK}$ denotes the Lie derivative, $\star$ means the Fr{\'e}chet derivative and $\dagger$ means the formal conjugation 
of the difference operator.
\end{Def}

If a symmetry of equation (\ref{ddeq}) is explicitly dependent on $\bu_i$ with $i\ne 0$, we say it is a generalised symmetry. 
An equation is integrable if it possesses infinitely many generalised symmetries
depending on finite sets of variables $\bu_i$, whose sizes are increasing. 
Symmetries  of an integrable systems can be generated by a recursion operator 
which is often nonlocal.
Sufficient conditions on nonlocal pseudo-difference recursion  operators which guarantee 
the production of an infinite hierarchy of commuting local symmetries have been discussed in \cite{mwx2}. 
In the list, we also give partial results on master symmetries (see \cite{mr97j:22044, mr2000e:37104} for
more details on master symmetries).

We now look at how the recursion, Hamiltonian and symplectic operators change under  transformations (difference substitutions).
If an evolutionary difference equation  (\ref{ddeq}) is connected by a difference substitution $\bu=\bF[\bv],\ \bF[\bv]\in\cF$ 
with another equation of the form 
\begin{eqnarray}\label{eqv}
 \bv_t=\bG[\bv]
\end{eqnarray}
then  recursion, 
Hamiltonian and symplectic operators ($\hat \cR, \hat \cH$ and $\hat\J$)
for the equation (\ref{eqv}) can be expressed in terms of  the corresponding operators ($\cR, \cH$ and $\J$) 
of the equation    (\ref{ddeq})
\begin{eqnarray}\label{trar}
\hat \cR=\bF_\star^{-1} \circ \cR\lvert_{\bu=\bF[\bv]} \circ \bF_\star, \qquad 
\hat\cH=\bF_\star^{-1} \circ \cH\lvert_{\bu=\bF[\bv]} \circ {\bF_\star^{\dagger}}^{-1}, \qquad
\hat\J=\bF_\star^{\dagger} \circ \J\lvert_{\bu=\bF[\bv]}\circ \bF_\star
\end{eqnarray}
where $\circ$ denotes the composition of the operators.

\begin{Ex} It is known that the Volterra equation $u_t=u(u_1-u_{-1})$  (section \ref{vol}) can be related to the modified 
Volterra equation  
\begin{equation}\label{modvol0}
 v_t=v^2 (v_1-v_{-1})
\end{equation}
 by a difference substitution (a Miura type transformation) $u=F[v]=v v_1$. 
Operator $\cH_1=u(\cS-\cS^{-1})u$ is a Hamiltonian operator for the Volterra equation. 
Notice that the Fr{\'e}chet derivative for $v v_1$ is $F_{\star}=v_1+v \cS$, and we have 
$F_\star^{-1} u=v (1+\cS)^{-1}$ and $u {F_\star^{\dagger}}^{-1}= \cS (1+\cS)^{-1} v$. Now we can find the Hamiltonian
operator $\hat{\cH}_1$ for the modified Volterra equation (\ref{modvol0})
\begin{eqnarray*}
&&\hat{\cH}_1=(v_1+v \cS)^{-1} v v_1 \left(\cS-\cS^{-1}\right) v v_1 (v_1+v \cS)^{-1}\\
&&\qquad=v (1+\cS)^{-1} (\cS^2-1) (\cS+1)^{-1} v\\
&&\qquad=v (\cS-1) (\cS+1)^{-1} v
\end{eqnarray*}
In the same way, we can compute the second Hamiltonian operator $\hat \cH_2$ and the recursion operator $\hat{\cR}$ 
for the modified Volterra equation (compare with section \ref{mvol}).
\end{Ex}

In Section \ref{Sec2} we discussed that the compatibility of a Darboux map (\ref{SPsi}) with the Lax operator (\ref{lax1})
\begin{eqnarray}\label{sysm}
 \cS (\Phi)=M \Phi, \qquad D_t (\Phi)=U(\bu; \lambda) \Phi
\end{eqnarray}
yields (\ref{backlundx})
\begin{equation}\label{zerocd}
D_t(M)=\cS(U) \ M-M\ U ,
\end{equation}
which is equivalent to an integrable system of differential-difference equations. 
In literature the compatibility condition (\ref{zerocd}) is often called a {\em zero curvature representation} 
or {\em Lax representation} of equation (\ref{ddeq}). Since it involves a Darboux matrix and a Lax operator, it is
more appropriate to call it a {\em Darboux-Lax representation}. However, we still call it a {\em Lax representation}
to be consistent with the literature.
In the following list, we simply give the expressions of both matrices $M$ and $U$ for Lax representations.

There are many publications  where an integrable system emerges as a compatibility condition of two linear problems 
with scalar linear difference operators $L$ and $A$
\begin{eqnarray}\label{syss}
L \phi=\lambda \phi, \qquad \phi_t=A \phi,
\end{eqnarray}
where $\phi$ is an eigenfunction of  $L$ corresponding to eigenvalue $\lambda$ and $\lambda_t=0$. 
Equation (\ref{ddeq}) is equivalent to their compatibility condition 
\begin{equation}\label{zerocs}
D_t(L)=[A,\ L]=A \ L-L\ A .
\end{equation}
This approach really resembles the original Lax formulation where differential operators are just replaced by difference ones.
In the theory of ordinary differential equations a scalar high order equation can be represented in the form
of a system of first order equations. Similar we can do in the case of scalar difference equations of high order 
and represent them by a first order difference systems. In this way we can re-write a scalar representation (\ref{syss}), (\ref{zerocs})
in the form of a first order matrix Darboux-Lax representation (\ref{sysm}), (\ref{zerocd}).

\begin{Ex}
For the Volterra Chain listed in section \ref{vol}, its matrix Lax representation is obtained from the scalar Lax representation
with $L= \cS +u \cS^{-1}$ and $A= \cS^2+u_1+u .$

Let $\Phi=(\phi^1,\ \phi^2)^T=(\phi,\ -\cS^{-1} \phi)^T$. We can rewrite  $L \phi=\lambda \phi$ in (\ref{syss}) as
\begin{eqnarray*}
 \cS(\Phi)=\cS \left(\begin{array}{c}\phi\\-\phi_{-1} \end{array}\right)=\left(\begin{array}{c} \cS \phi\\-\phi\end{array}\right)
= \left( \begin{array}{cc}\lambda & u \\ -1 & 0 \end{array}\right)
 \left(\begin{array}{c}\phi\\-\phi_{-1} \end{array}\right),
\end{eqnarray*}
that is, $\cS(\phi^1)=\lambda \phi^1+ u \phi^2$ and $\cS(\phi^2)=-\phi^1$. Now $\phi_t=A \phi$ can be rewritten as
\begin{eqnarray*}
&&\phi^1_t=\cS^2 \phi^1+(u_1+u) \phi^1=\cS (\lambda \phi^1+ u \phi^2)+(u_1+u) \phi^1
=\lambda \cS \phi^1+u \phi^1=(\lambda^2+u) \phi^1+\lambda u \phi^2
\end{eqnarray*}
and
$$\phi^2_t=-\cS^{-1} \phi^1_t=-(\lambda \phi^1+ u \phi^2+(u_{-1}+u) \cS^{-1}\phi^1)=-\lambda \phi^1+u_{-1} \phi^2 .$$
Thus
\begin{eqnarray*}
D_t (\Phi)=\left(\begin{array}{c}\phi^1_t\\\phi_t^{2} \end{array}\right)=\left( \begin{array}{cc} \lambda^2+u & \lambda u \\ -\lambda & u_{-1} \end{array}\right)
 \left(\begin{array}{c}\phi^1\\ \phi^{2} \end{array}\right).
\end{eqnarray*}
\end{Ex}
In our list we present either scalar or matrix Lax representation.

The reader should bear in mind that
\begin{itemize}
\item Relation between nonlocal difference operators, e.g. (\ref{relation}), should be understood as  identities in the non-commutative field of pseudo-difference Laurent series.
\item In the computations with pseudo-difference operators, we often use identities similar to ``integration by parts'':
\begin{eqnarray*}
 (\cS-1)^{-1} (f_1-f) (\cS-1)^{-1}=f (\cS-1)^{-1} -(\cS-1)^{-1} f_1
\end{eqnarray*}
\item Since for any constant $c\in \bbbc$ we have $(\cS-1) c=0$, the action of operators involving $(\cS-1)^{-1}$ is not uniquely defined. The corresponding results given in the list are up to these
``integration constants''.
\end{itemize}


\subsection{The Volterra Chain }\label{vol}
\begin{itemize}
\item  Equation \cite{Volterra}:
\begin{eqnarray}\label{voleq}
u_{t}=u(u_{1}-u_{-1})
\end{eqnarray}

\item Hamiltonian structure \cite{mr90g:58048, mr96h:35232}: $u_{t}=H_{i} \delta_u f_{i}$
\begin{eqnarray*}
&&\cH_{1}=u(\cE-\cE^{-1})u,\qquad f_{1}=u\\&&
\cH_{2}=u(\cE u\cE+u \cE+\cE u-u\cE^{-1}-\cE^{-1} u-\cE^{-1} u\cE^{-1})u,\qquad f_{2}=\frac{1}{2}\ln u
\end{eqnarray*}
\item  Recursion operator:
\begin{eqnarray*}
&&\cR=\cH_2 \cH_1^{-1}=u\cE+u+u_1+u \cE^{-1}+u(u_1-u_{-1})(\cE-1)^{-1}\frac{1}{u}\\
&&\qquad = u (\cE-\cE^{-1}) u \left( \frac{1}{u} (\cE-1)^{-1}+ \cE (\cE-1)^{-1} \frac{1}{u} \right)
\end{eqnarray*}
\item Non-trivial symmetry \cite{mr96h:35232, Yami}:
\begin{eqnarray*}
\cR (u_t)=u \left(u_{1} u_{2}+u_{1}^2+u u_1-u u_{-1}-u_{-1}^2- u_{-1} u_{-2}\right)
\end{eqnarray*}
\item Master symmetry \cite{mr96h:35232, mr97j:22044,levi2}:
\begin{eqnarray*}
\cR(u)=n u_t+u (2 u_1+u+u_{-1})
\end{eqnarray*}
\item Lax representation \cite{Bog88}:
\begin{eqnarray*}
L= \cS +u \cS^{-1} , \qquad A= \cS^2+u_1+u
\end{eqnarray*}
It can also be written in the matrix form
\begin{eqnarray}
M= \left( \begin{array}{cc}\lambda & u \\ -1 & 0 \end{array}\right); \qquad 
U= \left( \begin{array}{cc} \lambda^2+u & \lambda u \\ -\lambda & u_{-1} \end{array}\right)\label{volLax}
\end{eqnarray}
\end{itemize}
This equation is also known as Lotka-Volterra model, the Kac-van Moerbeke lattice or the Langmuir lattice \cite{manakov}. 
The so-called  Kac-Moerbeke-Langmuir equation \cite{mr90g:58048}
\begin{eqnarray*}
w_{\tau}= w (w_{1}^{\epsilon}-w_{-1}^{\epsilon}), \qquad \epsilon\neq 0 \ \mbox{is a constant}.
\end{eqnarray*}
is related to (\ref{voleq}) by the point transformation $u=w^\epsilon$ and $t=\epsilon \tau$. 
This equation is also written as
\begin{eqnarray*}
 w_t=\exp(w+w_1)-\exp(w+w_{-1}),
\end{eqnarray*}
which can be transformed into (\ref{voleq}) by the transformation $u=\exp(w+w_1)$.

\subsection{Modified Volterra equation }\label{mvol}
\begin{itemize}
\item Equation \cite{hirota730, Yami}:
$$u_{t}=u^{2}(u_{1}-u_{-1})$$

\item Hamiltonian structure \cite{tsuchida2,Yami}:
\begin{eqnarray*}
&&\cH_1=u (\cE-1) (\cE+1)^{-1} u, \qquad f_1=u u_1\\
&&\cH_2=u^2 (\cE-\cE^{-1}) u^2,\qquad f_2=\ln u
\end{eqnarray*}
\item Recursion operator:
\begin{eqnarray*}
\cR=\cH_2 \cH_1^{-1}=u^2 \cE+2uu_1+u^2\cE^{-1}+2 u^2(u_1-u_{-1})(\cE-1)^{-1}\frac{1}{u}
\end{eqnarray*}
\item Non-trivial symmetry \cite{Yami}:
\begin{eqnarray*}
\cR(u_t)= u^2 u_1^2 (u_2+u)-u^2 u_{-1}^2 (u+u_{-2})
\end{eqnarray*}
\item Master symmetry \cite{Yami}:
\begin{eqnarray*}
\cR(\frac{u}{2})=n u_t+\frac{u^2}{2} (3 u_1+u_{-1})
\end{eqnarray*}
\item Lax representation \cite{adler_dis}:
\begin{eqnarray*}
M= \left( \begin{array}{cc}0 & u \\ -u & \lambda \end{array}\right); \qquad 
U= \left( \begin{array}{cc} uu_{-1} & \lambda u_{-1} \\ -\lambda  u& \lambda^2+uu_{-1}\end{array}\right)
\end{eqnarray*}
\end{itemize}
The Modified Volterra equation is also known as discrete modified Korteweg-de Vries equation. 
Under the Miura transformation $w=u u_1$ it can be transformed into the Volterra Chain $w_t=w(w_1- w_{-1})$ as in section \ref{vol}.

\subsection{Yamilov's discretisation of the Krichever-Novikov equation}
\begin{itemize}
\item Equation \cite{Yami1}:
\begin{equation*}
u_{t}=\frac{R(u_1,u,u_{-1})}{u_1-u_{-1}}:=K^{(1)},
\end{equation*}
where $R$ is the polynomial with constant coefficients $\alpha, \beta, \gamma, \delta, \epsilon$ defined by
\begin{eqnarray}
&& R(u,v,w)=(\alpha v^2+2 \beta v+\gamma)uw+(\beta v^2+\lambda v+\delta) (u+w)+\gamma v^2+2 \delta v+\epsilon \ .\label{polyr}
\end{eqnarray}
\item Two non-trivial symmetries \cite{TTX, X, mr89k:58132, mwx1, mwx2, MW2011}:
\begin{eqnarray*}
&&K^{(2)} = \frac{R(u, u_{-1},u) R(u_{1},u, u_{1})}{(u_1-u_{-1})^2} \left(\frac{1}{u_2-u}+\frac{1}{u-u_{-2}} \right); \label{K2} \\
&&K^{(3)}= \frac{R(u_{1},u, u_{1}) R(u, u_{-1},u)}{(u_1-u_{-1})^2}\! \left(\!\frac{\cS^2 K^{(1)}}{(u_2-u)^2}\!+\!\frac{\cS^{-2} K^{(1)}}{(u-u_{-2})^2}\!\! \right) 
\!+ K^{(1)} K^{(2)}\! \left(\frac{1}{u_2-u}\!+\!\frac{1}{u-u_{-2}} \right)\!\label{K3}
\end{eqnarray*}
\item Hamiltonian structure \cite{mwx2, MW2011}:
\begin{eqnarray*}
&&{\cal{H}}= A\ \cS-\cS^{-1}\ A+\, 2\,K^{(1)}\,  (\cS-1)^{-1}{\cal{S}}  K^{(2)}\, +\, 2\, K^{(2)}\, (\cS-1)^{-1} K^{(1)};  \\
&&{\hat {\cal{H}}} = {\hat A}\, {\cal{S}}^2\,-\,
{\cal{S}}^{-2} {\hat A} +{\hat B} \cS-\cS^{-1} {\hat B} +\, K^{(2)}\,  (\cS-1)^{-1}({\cal{S}}+1)  K^{(2)}  \\
&&\qquad  +\, 2\,K^{(1)}\,  (\cS-1)^{-1}{\cal{S}}  K^{(3)}\, +\, 2\, K^{(3)}\,
(\cS-1)^{-1}  K^{(1)} \,, 
\end{eqnarray*}
where
\begin{eqnarray*}
&&A= \frac{R(u_{2},u_1, u_{2}) R(u_{1},u, u_{1}) R(u, u_{-1},u)}{(u_1-u_{-1})^2(u_2-u)^2} ;\\
&&{\hat A}=  \frac{R(u_{3},u_2, u_{3}) R(u_{2},u_1, u_{2}) R(u_{1},u, u_{1}) R(u, u_{-1},u)}{(u_1-u_{-1})^2(u_2-u)^2 (u_3-u_1)^2}; \\
&&{\hat B}=2 A \left(\frac{K^{(1)}}{u-u_{-2}} -\frac{\partial_{u} R(u_1,u, u_1)}{2(u_1-u_{-1})}
 +\frac{\partial^2  R(u_1,u, u_1)}{4 \partial {u} \partial {u_{1}}}  \right) + \frac{2 R(u, u_{-1},u)}{(u_1-u_{-1})^2} \cS \left( K^{(1)} K^{(2)}\right)
\end{eqnarray*}
\item Symplectic operator: 
\begin{eqnarray*}
\J&=& \frac{1}{R(u_{1},u, u_{1})}\ \cS -\cS^{-1}\ \frac{1}{R(u_{1},u, u_{1})} 
\end{eqnarray*}
\item Recursion operator  \cite{mwx2, MW2011}: 
\begin{eqnarray*}
\cR=\cH \J \quad \mbox{and} \quad {\hat \cR}={\hat {\cal{H}}} \J
\end{eqnarray*}
\end{itemize}
The recursion operators $\cR$ and
$\hat{\cR}$  satisfy the algebraic equation
\begin{eqnarray}
 (2 {\hat \cR} -I_3)^2=4 (\cR+I_2)^3- g_2 (\cR+I_2)-g_3, \label{relation}
\end{eqnarray}
where $I_2, I_3, g_2$ and $g_3$ are the relative and modular invariants
related to $h=R(u_1,u, u_1)$
and a quartic polynomial $f(u)= (\partial_{u_{1}} h)^2
-2 h \partial_{u_{1}}^2 h $ defined by
\begin{eqnarray*}
&&g_2=\frac{1}{48}\left(2 f f^{IV}-2 f'f'''+(f'')^2\right)\, ,\\
&&g_3=\frac{1}{3456}\left( 12 f f''f^{IV}-9 (f')^2 f^{IV}-6 f (f''')^2+6 f' f''
f''' -2 (f'')^3\right)\, ,\\
&&I_2=\frac{1}{6} \left(h \partial_u^2\partial_{u_{1}}^2 h-(\partial_u h)
(\partial_u \partial_{u_{1}}^2 h)-(\partial_{u_{1}} h) (\partial_u^2
\partial_{u_{1}} h)
+(\partial_u^2 h) (\partial_{u_{1}}^2 h)\right)+\frac{1}{12}(\partial_u \partial_{u_{1}} h)^2\, ,\\
&&I_3=\frac{1}{4} \det \left(\begin{array}{ccc}\ h\ &\ \partial_u h\ &\
\partial_u^2 h\ \\
\partial_{u_{1}} h & \partial_{u_{1}} \partial_u h & \partial_{u_{1}}
\partial_u^2 h\\
\partial_{u_{1}}^2 h & \partial_{u_{1}}^2 \partial_u h &
\partial_{u_{1}}^2 \partial_u^2 h
\end{array}\right)\, .
\end{eqnarray*}

\subsection{Integrable Volterra type equations}\label{yamivol}
The classification of integrable Volterra type equations of the form
$$u_t=f(u_{-1},u,u_{1}),$$
where $f$ is a smooth function of all its variables was obtained by Yamilov using the symmetry approach. In his remarkable review paper \cite{Yami}, he presented the following complete list
of integrable Volterra-type equations (with higher order conservation laws) up to the point transformations:
\begin{eqnarray}
&&\mbox{V1}. \qquad  u_t=P(u) (u_1-u_{-1}) \label{v1}\\
&&\mbox{V2}. \qquad u_t=P(u^2) \left(\frac{1}{u_1+u}-\frac{1}{u+u_{-1}}\right) \label{v2}\\
&&\mbox{V3}. \qquad u_t=Q(u) \left(\frac{1}{u_1-u}+\frac{1}{u-u_{-1}}\right)\label{v3}\\
&&\mbox{V4}. \qquad u_t=\frac{R(u_1,u,u_{-1})+\nu R(u_1,u, u_1)^{1/2} R(u_{-1},u,u_{-1})^{1/2}}{u_1-u_{-1}},\qquad \nu\in \{0,\pm1\}\label{v4}\\
&&\mbox{V5}. \qquad u_t=y(u_1-u)+y(u-u_{-1}),\qquad y'=P(y)\label{v5}\\
&&\mbox{V6}. \qquad u_t=y(u_1-u)\ y(u-u_{-1})+\mu,\qquad y'=\frac{P(y)}{y}, \quad \mu\in \cC\label{v6}\\
&&\mbox{V7}. \qquad u_t=\frac{1}{y(u_1-u)+y(u-u_{-1})}+\mu,\qquad y'=P(y^2) \quad \mu\in \cC\label{v7}\\
&&\mbox{V8}. \qquad u_t=\frac{1}{y(u_1+u)-y(u+u_{-1})},\qquad y'=Q(y)\label{v8}\\
&&\mbox{V9}. \qquad u_t=\frac{y(u_1+u)-y(u+u_{-1})}{y(u_1+u)+y(u+u_{-1})},\qquad y'=\frac{P(y^2)}{y}\label{v9}\\
&&\mbox{V10}. \qquad u_t=\frac{y(u_1+u)+y(u+u_{-1})}{y(u_1+u)-y(u+u_{-1})},\qquad y'=\frac{Q(y)}{y}\label{v10}\\
&&\mbox{V11}. \qquad u_t=\frac{(1-y(u_1-u))(1-y(u-u_{-1}))}{y(u_1-u)+y(u-u_{-1})}+\mu,\qquad y'=\frac{P(y^2)}{1-y^2}, \quad \mu\in \cC,\label{v11}
\end{eqnarray}
where $P$ and $Q$ are polynomials with constant coefficients $\alpha, \beta, \gamma, \delta, \epsilon$ defined by
\begin{eqnarray}
&&P(u)=\alpha u^2 +\beta u +\gamma ,\label{polyp}\\ 
&& Q(u)=\alpha u^4 +\beta u^3 +\gamma u^2 +\delta u +\epsilon,\label{polyq}
\end{eqnarray}
and the polynomial $R$ is defined by (\ref{polyr}).
As stated in the paper \cite{Yami}, the problem of constructing the generalized symmetries for all equations (V1)--(V11) 
remains open although the master symmetries for some forms of equations in the list are known \cite{mr96h:35232, mr97j:22044}.
We know that the Miura transformation $\tilde u=y(u_{1}-u)$ transforms equations (V5) and (V6) to (V1) and equations (V7) and (V11) into (V2), and
the Miura transformation $\tilde u=y(u_{1}+u)$ transforms equation (V9) to (V2) and equations (V8) and (V10) into (V3)
\cite{Yami}.
In the follows, we present the recursion operators,  Hamiltonian operators and master symmetries for the first four equations.
The corresponding operators for other equations can be obtained via the Miura transformations.

\subsubsection{V1 equation (\ref{v1})}
\begin{itemize}
\item Hamiltonian structure:
$$\cH=P(u) (\cE-\cE^{-1}) P(u)
$$
\item Symplectic operator:
$$
\alpha (\cE-\cE^{-1}) +(\alpha u_1+\beta +\alpha u_{-1}) \cS (\cS-1)^{-1} \frac{P'(u)}{P(u)} 
+ \frac{P'(u)}{P(u)} (\cS-1)^{-1} (\alpha u_1+\beta +\alpha u_{-1})
$$
\item Recursion operator:
\begin{eqnarray*}
\cR=P(u) \cS +2 \alpha u u_1 +\beta (u+u_1) +P(u) \cS^{-1} + u_t (\cS-1)^{-1} \frac{P'(u)}{P(u)} 
\end{eqnarray*}
\item Non-trivial symmetry :
\begin{eqnarray*}
\cR(u_t)=P(u) \left(P(u_1) u_2+\alpha u u_1^2+\beta (u +u_1) u_1-P(u_{-1}) u_{-2}-\alpha u u_{-1}^2-\beta (u +u_{-1}) u_{-1}\right)
\end{eqnarray*}
\item Master symmetry:
\begin{eqnarray*}
&&n u_t + P(u) (c u_1+\frac{\beta}{\alpha}+(2-c) u_{-1}),\ \  c\in \cC, \quad \mbox{when $\alpha\neq 0$}; \\ 
&&n u_t +P(u) (c u_1+u+(3-c) u_{-1}), \ \ c\in \cC, \quad \mbox{when $\alpha=0$}
\end{eqnarray*}
\item Lax representation:
\begin{itemize}
\item The case $\alpha=\beta=0$ is a linear equation.
\item The case $\alpha=0,\ \beta\ne0$ reduces to the Volterra equation (\ref{voleq}) by linear substitution  $u\mapsto \beta^{-1} (u-\gamma)$. 
Thus it has the Lax representation (\ref{volLax}).
\item In the case $\alpha\ne0$ the following linear substitution $t\mapsto \alpha^{-1} t$ and $u\mapsto  u-\beta/2$ transforms the polynomial $P(u)$ in {\bf V1} to the 
form $P(u)=u^2+c$, where $c=\gamma/\alpha-\beta^2/(4\alpha^2)$. The corresponding Lax representation is of the form \cite{adler_dis}
\begin{eqnarray*}
M= \left( \begin{array}{cc}c\lambda^{-1} & u \\ -u & \lambda \end{array}\right); \qquad 
U= \left( \begin{array}{cc} c^2 \lambda^{-2}+uu_{-1} & c\lambda^{-1} u+\lambda u_{-1} \\-c\lambda^{-1}u_{-1} -\lambda  u& \lambda^2+uu_{-1}\end{array}\right)
\end{eqnarray*}
\end{itemize}
\end{itemize}
This equation includes both Volterra chain in section \ref{vol} and modified Volterra equation in section \ref{mvol}. 
For nonlinear equation, that is, $\alpha \beta \neq 0$, its Hamiltonian $f$ depends on the coefficients in polynomial 
$P$ defined by (\ref{polyp}). When $\alpha \neq 0$, we
take $f=\frac{1}{2 \alpha}\ln P(u)$, otherwise, we take $f=\frac{u}{\beta}$.

Its Hamiltonian operator, symplectic operator and recursion operator satisfy the following unexpected relation
$$ \cH \J=\alpha \cR^2+\beta^2 \cR+2 \gamma (\beta^2 -2 \alpha \gamma) .
$$

\subsubsection{V2 equation (\ref{v2})}\label{secv2}
\begin{itemize}
\item Hamiltonian structure:
$$\cH=\frac{P(u^2)}{u_1+u} \cS \frac{P(u^2)}{u+u_{-1}}-\frac{P(u^2)}{u+u_{-1}} \cS^{-1}\frac{P(u^2)}{u_1+u}
-u_t (\cS+1)(\cS-1)^{-1} u_t
$$
\item Symplectic operator:
\begin{eqnarray*}
&&\J=\frac{1}{(u+u_1)^2} \cS-\cS^{-1} \frac{1}{(u+u_1)^2}+ \delta_u \rho (\cS+1) (\cS-1)^{-1} \delta_u \rho\\
&&\quad -(\beta^2-4 \alpha \gamma) \frac{u}{P(u^2)} (\cS-1)(\cS+1)^{-1} \frac{u}{P(u^2)},
\end{eqnarray*}
where $\rho=\frac{1}{2} \ln \frac{(u+u_1)^2}{P(u^2)}$ and thus 
\begin{eqnarray*}
 \delta_u \rho=\frac{1}{u_1+u}+\frac{1}{u+u_{-1}}
- \frac{{P(u^2)}'}{2 P(u^2)}, \qquad  {P(u^2)}'=2u(2 \alpha u^2+\beta)
\end{eqnarray*}
\item Recursion operator:
\begin{eqnarray*}
&&\cR=\frac{P(u^2)}{(u_1+u)^2} \cS +P(u^2) \left(\frac{1}{(u+u_1)^2}+\frac{2}{(u+u_1)(u+u_{_1})}-\frac{1}{(u+u_{-1})^2}\right)\\
&&\quad+{P(u^2)}'\left(\frac{1}{2u}- \frac{1}{u+u_1}\right)
 +\frac{P(u^2)}{(u+u_{-1})^2} \cS^{-1}+2 u_t (\cS-1)^{-1} \delta_u \rho
\end{eqnarray*}
\item Non-trivial symmetry :
\begin{eqnarray*}
&&\cR(u_t)=\frac{P(u^2) P(u_1^2)}{(u+u_1)^2} \left(\frac{1}{u_1+u_2}-\frac{1}{u_1-u_{-1}}\right)
-\frac{P(u^2) P(u^2_{-1})}{(u+u_{-1})^2}\left(\frac{1}{u_{-1}+u_{-2}}-\frac{1}{u_{-1}-u_1}\right)\\
&&\qquad\quad +2 P(u^2) \frac{\alpha u_1^2u_{-1}^2+\beta u_1 u_{-1}+\gamma}{(u+u_1)(u_1-u_{-1})(u+u_{-1})}
\end{eqnarray*}
\item Master symmetry:
\begin{eqnarray*}
&&n u_t + \frac{P(u^2)}{u+u_{-1}} -\alpha u^3-\beta u, \quad \mbox{when $\gamma= 0$}, \\ 
&&n u_t + \frac{P(u^2)}{u+u_{-1}} -\alpha u^3-\frac{\beta}{2} u, \quad \mbox{when $\beta^2-4 \alpha\gamma= 0$}, \\ 
&&n u_t +\frac{P(u^2)}{u+u_{-1}},  \quad \mbox{when $\alpha=0$}.
\end{eqnarray*}
\end{itemize}
Here we only found the master symmetries for some special cases.  For the given Hamiltonian operator above, we
 have $\cH \delta_u \rho=\cR (u_t)$. Besides, the Hamiltonian operator, symplectic operator and
 the recursion operator satisfy the following relation: 
$$\cH \J=\cR^2-2 \beta\ \cR +\beta^2-4 \alpha \gamma.$$

The Calogero-Degasperis Lattice \cite{mr97j:22044}
\begin{eqnarray*}
u_{t}=\frac{1}{4}(1-u^2)(b^2-a^2 u^2)(\frac{1}{u_{1}+u}-\frac{1}{u+u_{-1}})
\end{eqnarray*}
is one of special cases of the V2 equation.  The authors of \cite{mr97j:22044} presented its master symmetry 
by introducing the time dependence for the coefficients $a$ and $b$, which is in different form from what we did. 

\subsubsection{V3 equation (\ref{v3})}
\begin{itemize}
\item Hamiltonian structure:
$$\cH=\frac{Q(u)}{u_1-u} \cS \frac{Q(u)}{u-u_{-1}}-\frac{Q(u)}{u-u_{-1}} \cS^{-1}\frac{Q(u)}{u_1-u}
+u_t (\cS+1)(\cS-1)^{-1} u_t
$$
\item Symplectic operator:
\begin{eqnarray*}
&&\J=\frac{1}{(u_1-u)^2} \cS-\cS^{-1} \frac{1}{(u_1-u)^2}- \delta_u \rho (\cS+1) (\cS-1)^{-1} \delta_u \rho\\
&&\quad-(2 \alpha \gamma -\beta^2/4) \frac{u^2}{Q(u)} (\cS-1) (\cS+1)^{-1} \frac{u^2}{Q(u)}
-\beta \delta \frac{u}{Q(u)} (\cS-1)(\cS+1)^{-1} \frac{u}{Q(u)}\\
&&\quad-(2 \gamma \epsilon -\delta^2/4) \frac{1}{Q(u)} (\cS-1)(\cS+1)^{-1} \frac{1}{Q(u)}\\&&\quad 
+(2 \alpha \delta+\beta \gamma) (\frac{u^2}{Q(u)}  (\cS+1)^{-1} \frac{u}{Q(u)} -\frac{u}{Q(u)} \cS  (\cS+1)^{-1} \frac{u^2}{Q(u)})
\\&&\quad 
-(\beta \delta/2- \gamma^2-4 \alpha \epsilon) (\frac{u^2}{Q(u)}  (\cS+1)^{-1} \frac{1}{Q(u)} -\frac{1}{Q(u)} \cS  (\cS+1)^{-1} \frac{u^2}{Q(u)})
\\&&\quad 
+(\gamma \delta+2 \beta \epsilon) (\frac{u}{Q(u)}  (\cS+1)^{-1} \frac{1}{Q(u)} -\frac{1}{Q(u)} \cS  (\cS+1)^{-1} \frac{u}{Q(u)}),
\end{eqnarray*}
where  $\rho=\frac{1}{2} \ln \frac{Q(u))}{(u_1-u)^2}$ and thus 
\begin{eqnarray*}
 \delta_u \rho=\frac{1}{u_1-u}-\frac{1}{u-u_{-1}}+ \frac{Q'(u)}{2 Q(u)}
\end{eqnarray*}
\item Recursion operator:
\begin{eqnarray*}
&&\cR=\frac{Q(u)}{(u_1-u)^2} \cS +Q(u) \left(\frac{1}{(u-u_{-1})^2}+\frac{2}{(u_1-u)(u-u_{_1})}-\frac{1}{(u_1-u)^2}\right)\\
&&\quad-\frac{Q'(u)}{u_1-u}-2 \alpha u^2-\beta u -\gamma+\frac{Q(u)}{(u-u_{-1})^2} \cS^{-1} - 2 u_t (\cS-1)^{-1}\delta_u \rho
\end{eqnarray*}
\item Non-trivial symmetry :
\begin{eqnarray*}
&&\cR(u_t)=\frac{Q(u) Q(u_1)}{(u_1-u)^2(u_2-u_1)} +\frac{Q(u) Q(u_{-1})}{(u-u_{-1})^2(u_{-1}-u_{-2})}+\alpha Q(u) (u_1-u_{-1})\\
&&\qquad\quad +Q(u) \left(\frac{Q(u)}{(u_1-u)(u-u_{-1})}+2 \alpha u^2+\beta u\right)\left(\frac{1}{u_{1}-u}+\frac{1}{u-u_{-1}}\right)
\end{eqnarray*}
\item Master symmetry:
\begin{eqnarray*}
&&n u_t - \frac{Q(u)}{u+u_{-1}} +\alpha u^3+\beta u^2+\gamma u, \quad \mbox{when $\delta=\epsilon= 0$}, \\ 
&&n u_t - \frac{Q(u)}{u+u_{-1}}, \quad \mbox{when $ \alpha= \beta=0$}
\end{eqnarray*}
\end{itemize}
Similar to V2 equation in section \ref{secv2}, we have not found the master symmetry valid for the polynomial $Q$ 
(\ref{polyq}) with arbitrary coefficients. However, if we look for the master symmetry of the form
\begin{eqnarray*}
 n u_t - \frac{Q(u)}{u+u_{-1}} +\sum_{i=0}^3 c_i u^i, \qquad c_i\in \cC .
\end{eqnarray*}
we can determine the constants $c_i$ for certain polynomial $Q$. Two examples are listed above.

Notice that $\cH \delta_u \rho=\cR (u_t)$ and the product of 
the Hamiltonian operator and symplectic operator gives rise of the square of the recursion operator,
that is, $$\cH \J=\cR^2. $$


\subsection{The Narita-Itoh-Bogoyavlensky lattice}
\begin{itemize}
\item Equation \cite{Bog88, narita, itoh}:
$$u_t=u (\sum_{k=1}^p u_{k}-\sum_{k=1}^p u_{-k}), \qquad p\in\mathbb{N}$$
\item Hamiltonian structure \cite{mr93c:58096}:
\begin{eqnarray*}
\cH=u(\sum_{i=1}^{p} \cS^i-\sum_{i=1}^{p} \cS^{-i}) u, \qquad f= u
\end{eqnarray*}
\item Recursion operator \cite{wang12}:
 \begin{eqnarray*}
\cR=u (\cS -\cS^{-p}) (\cS -1)^{-1} \prod_{i=1}^{\rightarrow{p}} (\cS^{p+1-i} u-u\cS^{-i} )
(\cS^{p-i} u-u \cS^{-i})^{-1} \ ,
\end{eqnarray*}
where the notation $\prod_{i=1}^{\rightarrow{p}}$ is denoted the order of the value $i$, from $1$ to $p$, that is,\\
$\prod_{i=1}^{\rightarrow{p}}a_i=a_1 a_2 \cdots a_p .$
\item Non-trivial symmetry:
\begin{eqnarray*}
\cR(u_t)=u (1-\cE^{-(p+1)}) \cE^{1-p} \sum_{0\leq i\leq j\leq 2p-1} u_j u_{i+p}
\end{eqnarray*}
\item Master symmetry \cite{wang12}: $\cR(u)$
\item Lax representation \cite{Bog88}:
\begin{eqnarray*}
L= \cS +u \cS^{-p} , \qquad A= (L^{(p+1)})_{\geq 0},
\end{eqnarray*}
where $\geq 0$ means taking the terms with non-negative power of $\cS$ in $L^{(p+1)}$.
\end{itemize}
For $p=1, 2$ or $3$, a few higher order symmetries are explicitly given in \cite{mr93c:58096}, where the authors
also studied their Hamiltonian operator, recursion operator and master symmetry for $p=1, 2$.

The Narita-Itoh-Bogoyavlensky lattice is known as an integrable
discretisation for the Korteweg-de Vries equation. It can also be presented as
\begin{eqnarray*}
&&v_t=v (\prod _{k=1}^p v_{k}-\prod_{k=1}^p v_{-k}) .
\end{eqnarray*}
which is related to the Narita-Itoh-Bogoyavlensky lattice via the
transformation $u= \prod_{k=0}^{p-1} v_{k}$ for fixed~$p$.

Taking $p=1$, we get the well-known Volterra chain in section \ref{vol}.
Thus they can be regarded as the generalisation of the Volterra chain.

Let $u=\prod_{k=0}^p w_{k}$. Then $w$ satisfies the so-called the modified Bogoyavlensky chain
\begin{eqnarray*}
w_t=w^2 (\prod_{k=1}^p w_{k}-\prod_{k=1}^p w_{-k}).
\end{eqnarray*}

The recursion operator given above for the Narita-Itoh-Bogoyavlensky lattice is highly nonlocal (so is the master symmetry). Recently, Svinin \cite{svin09} derived the explicit formulas for its generalised symmetries 
in terms of a family of homogeneous difference polynomials. The properties of 
these homogeneous difference polynomials \cite{svin11} enable us to prove the locality of 
its infinitely many symmetries \cite{wang12}.

A family of integrable lattice hierarchies associated with fractional
Lax operators was introduced by Adler and Postnikov \cite{adler1, adler2}. One simple
example is
\begin{eqnarray}\label{adler}
u_t=u^2 (\prod_{k=1}^p u_k -\prod_{k=1}^p u_{-k})-u (\prod_{k=1}^{p-1} u_k -\prod_{k=1}^{p-1} u_{-k}), \qquad 2\leq p\in\bbbn ,
\end{eqnarray}
which is an integrable discretisation for the Sawada-Kotera
equation.  It can be considered as inhomogeneous 
generalisation of the Bogoyavlensky type lattices. The problem to construct the Hamiltonian structure and recursion operator for such family of equations
is still open.

Even for in the scalar case,  the classification of higher order of integrable evolutionary differential-difference 
equations is still open.

\subsection{The Toda Lattice}\label{toda}
\begin{itemize}
\item Equation \cite{toda67}:
\begin{eqnarray*}
q_{tt}=\exp(q_{1}-q)-\exp(q-q_{-1})
\end{eqnarray*}
In the Manakov-Flaschka coordinates \cite{flaschka1, manakov} defined by $u=\exp(q_1-q),\  v=q_{t}$, it can be rewritten as two-component evolution system:
\begin{eqnarray}\label{todaeq}
\left\{ {\begin{array}{l} u_t= u(v_{1}-v)\\
   v_t=  u-u_{-1}  \end{array} } \right.
 \end{eqnarray}
\item Hamiltonian structure  \cite{mr90g:58048, mr97j:22044, kp85}:
\begin{eqnarray*}
 &&\cH_1=\left(\begin{array}{cc} 0 & u(\cE-1)  \\
 (1-\cE^{-1})u &0\end{array}\right), \qquad f_1=u+\frac{v^{2}}{2}\\
 &&\cH_{2}=\left( {\begin{array}{cc}
 u(\cE-\cE^{-1})u& u(\cE-1)v  \\
  v(1-\cE^{-1})u&u\cE-\cE^{-1}u
 \end{array} } \right),\qquad f_{2}=v
\end{eqnarray*}
\item Recursion operator:
\begin{eqnarray*}
&& \cR=\cH_{2}\cH_{1}^{-1}=\left( {\begin{array}{cc}
 v_1+u(v_1-v)(\cE-1)^{-1}\frac{1}{u}& u\cE+u  \\
  1+\cE^{-1}+(u-u_{-1})(\cE-1)^{-1}\frac{1}{u}&v
 \end{array} } \right)\\
 &&\quad=\left(\begin{array}{cc}  v_1& u\cE+u  \\
  1+\cE^{-1}&v \end{array}\right)+\left(\begin{array}{c} u(v_1-v) \\u-u_{-1} \end{array}\right) (\cS-1)^{-1}
 \left(\begin{array}{cc} \frac{1}{u} & 0 \end{array}\right)
\end{eqnarray*}
\item Non-trivial symmetry:
\begin{eqnarray*}
\cR\left(\begin{array}{c} u_t \\v_t \end{array}\right)=\left( \begin{array}{c}
 u(v_{1}^2-v^{2}+u_{1}-u_{-1})\\
    u(v_{1}+v)-u_{-1}(v_{-1}+v)
 \end{array}  \right)
\end{eqnarray*}
\item Master symmetry \cite{mr97j:22044}:
\begin{eqnarray*}
\cR\left(\begin{array}{c} u \\ \frac{v}{2} \end{array}\right)=\left( {\begin{array}{c}
 n u_t+\frac{3}{2}  u v_1+\frac{1}{2} u v \\
 n v_t + u+u_{-1}+\frac{v^2}{2}
 \end{array} } \right)
 \end{eqnarray*}
 \item Lax representation:
\begin{eqnarray*}
M= \left( \begin{array}{cc} \lambda+v_1&  u \\ -1 & 0 \end{array}\right); \qquad 
U= \left( \begin{array}{cc}0 &-u\\1&\lambda+v\end{array}\right)
\end{eqnarray*}
\end{itemize}
The Hirota nonlinear Lumped Network equation \cite{hirota73}
\begin{eqnarray*}
\left\{ {\begin{array}{l} u_t= v_1-v\\
   v_t= v(u-u_{-1})  \end{array} } \right.
 \end{eqnarray*}
is related to the Toda lattice (\ref{todaeq}) by a simple invertible transformation. Namely, let
$u=q$ and $v=p_{-1}$. Then the variables $p$ and $q$ satisfy the Toda equation. All its properties 
can be obtained via those for the Toda lattice.

\subsection{A Relativistic Toda system }\label{secrtoda}
 \begin{itemize}
 \item Equation \cite{ruijsenaars1}:
 \begin{eqnarray*}
q_{tt}=q_{t}q_{-1t}\frac{\exp(q_{-1}-q)}{1+\exp(q_{-1}-q)}-q_{t}q_{1t}\frac{\exp(q-q_{1})}{1+\exp(q-q_{1})}
\end{eqnarray*}
Let us introduce the following dependent variables \cite{suris1,mr90j:58061}
\begin{eqnarray*}
u=\frac{q_{t} \exp(q-q_1)}{1+\exp(q-q_1)},\qquad
v=\frac{q_{t}}{1+\exp(q-q_1)}.
\end{eqnarray*}
Then the equation can be written as 
\begin{eqnarray}\label{rtoda}
\left\{ {\begin{array}{l} u_t= u(u_{-1}-u_1+v-v_1)\\
   v_t= v(u_{-1}-u)  \end{array} } \right.
 \end{eqnarray}
\item Hamiltonian  structure \cite{mr90j:58061}:
\begin{eqnarray*}
&&\cH_1=\left( {\begin{array}{cc}
0 &  u(1-\cE) \\
 (\cE^{-1}-1)u & u \cE-\cE^{-1} u
 \end{array} } \right)
,\qquad f_1=\frac{1}{2}(u^2+v^2)+ uv +u_1 u+u v_1\\
&&\cH_2=\left( {\begin{array}{cc}
u(\cE^{-1}-\cE)u&  u(1-\cE)v \\
 v(\cE^{-1}-1)u &0
 \end{array} } \right)
,\qquad f_2=u+v
\end{eqnarray*}
\item Recursion operator \cite{mr90j:58061}:
\begin{eqnarray*}
&&\cR=\cH_{2}\cH_{1}^{-1}=\left( {\begin{array}{cc}
u\cE+u+v_{1}+u_{1}+u\cE^{-1}-u(v-v_1+u_{-1}-u_{1})(\cE-1)^{-1}\frac{1}{u} & u\cE+u \\
 v+v\cE^{-1}-v(u_{-1}-u)(\cE-1)^{-1}\frac{1}{u} & v
 \end{array} } \right) \\
&&\qquad = \left( {\begin{array}{cc}
u\cE+u+v_{1}+u_{1}+u\cE^{-1} & u\cE+u \\
 v+v\cE^{-1} & v
 \end{array} } \right)-\left(\begin{array}{c} u_t \\v_t \end{array}\right) (\cS-1)^{-1}
 \left(\begin{array}{cc} \frac{1}{u} & 0 \end{array}\right)
\end{eqnarray*}
\item Non-trivial symmetry :
\begin{eqnarray*}
\left(\!\!\! \begin{array}{c}
u u_{-1}(u+u_{-1}+u_{-2}+2v +v_{-1})-u u_{1}(u_{2}+u_{1}+u+2 v_1+v_2)
 +u^{2}( v-v_{1})+
 u(v^2-v_{1}^{2})\\
v u_{-1} \left( u_{-2}+u_{-1}+v+v_{-1}\right)-u v (u_1+u+v_1+v)
 \end{array} \!\!\!  \right) 
\end{eqnarray*}
\item Master symmetry \cite{mr90j:58061}:
\begin{eqnarray*}
\cR\left(\begin{array}{c} u \\ v \end{array}\right)=\left( {\begin{array}{c} -n u_t+u(v+2 v_1+u+2 u_1+u_{-1})\\
-n v_t+v(u+v+u_{-1})
\end{array} } \right), \quad \left( (\cS-1)^{-1} 1=n \right)
\end{eqnarray*}
 \item Lax representation:
\begin{eqnarray*}
M= \left( \begin{array}{cc} \lambda v-\lambda^{-1}&  u_{-1} \\ -1 & 0 \end{array}\right); \qquad 
U= \left( \begin{array}{cc}-\lambda^{-2}-u_{-1} &\lambda^{-1} u_{-1}\\ -\lambda^{-1}& -u_{-2}-v_{-1}\end{array}\right)
\end{eqnarray*}
\end{itemize}
As noticed in \cite{mr90j:58061}, the inverse of this recursion operator $\cR$ is also weakly nonlocal: 
\begin{eqnarray*}
&&\cR^{-1}=\cH_1 \cH_2^{-1}= \left( \begin{array}{cc}
\frac{1}{v_1} &-\frac{u}{v_1^2} \cE+\frac{u}{v^2}-\frac{2u}{v v_1}\\
-\cE^{-1} \frac{1}{v}-\frac{1}{v_1}& \frac{u}{v_1^2} \cE+\cE^{-1} \frac{u}{v^2} +\frac{2 u}{v v_1}+\frac{1}{v}\end{array}
\right)\\
&&\qquad+\left(\begin{array}{c} \frac{u}{v_{1}} -\frac{u}{v} \\\frac{u_{-1}}{v_{-1}}-\frac{u}{v_1} \end{array}\right) 
(\cS-1)^{-1} \left(\begin{array}{cc} \frac{1}{u} & -\frac{2}{v} \end{array}\right)
\end{eqnarray*}
However, recursion operators $\cR$ and $\cR^{-1}$ have different starting point, i.e., seeds. For $\cR$, it starts with acting on the right hand of the equation while
the seed for $\cR^{-1}$ is $\sigma=\left( \begin{array}{l}\frac{u}{v_{1}} -\frac{u}{v} \\
\frac{u_{-1}}{v_{-1}}-\frac{u}{v_1} \end{array}\right)$. Moreover, $\cR$ acting on $\sigma$ and $\cR^{-1}$ acting on the right hand of the equation do not give rise to new symmetries.

There is a Miura transformation $u=-\frac{u'_{-1}}{v' v'_{-1}}, v=-\frac{1}{v'_{-1}}$ 
between the flow corresponding to $\sigma$  and the equation, where $u'$ and $v'$ denote dependent variables for $\sigma$.

In \cite{suris1}, the author studied other integrable equations related the relativistic Toda lattice. For example, the equation
$$q_{tt}=q_{-1t}\exp(q_{-1}-q)-\exp(2q_{-1}-2q)-q_{1t} \exp(q-q_1)+\exp(2q-2q_1)
$$
can also be rewritten as system (\ref{rtoda}) by setting
$$ u=\exp(q-q_1), \qquad v=q_t-\exp(q_{-1}-q)-\exp(q-q_1) .
$$ 

\subsection{Two-component Volterra lattice }
\begin{itemize}
\item Equation \cite{mr90g:58048}:
\begin{eqnarray}\label{2vol}
\left\{ {\begin{array}{l} u_t= u(v_{1}-v)\\
   v_t=  v(u-u_{-1})  \end{array} } \right.
 \end{eqnarray}
\item Hamiltonian structure \cite{mr90g:58048}:
\begin{eqnarray*}
&&\cH_1=\left( {\begin{array}{cc}
 0&u(\cE-1)v  \\
  v(1-\cE^{-1})u&0
 \end{array} } \right),\qquad
f_1=u+v\\&&
\cH_2=\left( {\begin{array}{cc}
 u(\cE v-v\cE^{-1})u&u(u\cE-u+\cE v-v)v  \\
 v(u-\cE^{-1}u+v-v\cE^{-1})u &v(u\cE-\cE^{-1}u)v
 \end{array} } \right),\qquad f_2=\ln u
\end{eqnarray*}
\item Recursion operator:
\begin{eqnarray*}
&& \cR=\cH_{2}\cH_{1}^{-1}=\left( {\begin{array}{cc}
 u+v_{1}+u(v_1-v)(\cE-1)^{-1}\frac{1}{u}&u\cE+\frac{uv_{1}}{v}+u(v_1-v)(\cE-1)^{-1}\frac{1}{v}\vspace{1mm}  \\
v+v\cE^{-1}+v(u-u_{-1})(\cE-1)^{-1}\frac{1}{u}& u+v+v(u-u_{-1})(\cE-1)^{-1}\frac{1}{v}
 \end{array} } \right)\\
 &&\quad=\left(\begin{array}{cc}  u+v_{1}& u\cE+\frac{uv_{1}}{v}  \\
v+v\cE^{-1}&u+v \end{array}\right)+\left(\begin{array}{c} u(v_1-v) \\v (u-u_{-1}) \end{array}\right) (\cS-1)^{-1}
 \left(\begin{array}{cc} \frac{1}{u} & \frac{1}{v} \end{array}\right)
\end{eqnarray*}
\item Non-trivial symmetry:
\begin{eqnarray*}
\cR\left(\begin{array}{c} u_t \\v_t \end{array}\right)=\left( \begin{array}{c}
  {u}^{2} \left( v_{1}-v \right) +u \left( {v_{1}}^{2}-{v}^{2} + v_{1}u_{1}-vu_{-1} \right)\\
  {v}^{2} \left( u-u_{-1} \right) +v ( u^{2}-u_{-1}^{2}  + uv_{1}-u_{-1}v_{-1} )
 \end{array}  \right)
\end{eqnarray*}
\item Master symmetry \cite{mr97j:22044}:
\begin{eqnarray*}
\cR\left(\begin{array}{c} u \\ v \end{array}\right)=\left( {\begin{array}{c}
2 n u_t+u^2+3uv_1  \\
2 n v_t+vu_{-1}+2 u v+v^2
 \end{array} } \right)
 \end{eqnarray*}
\item Lax representation \cite{suris2}:
\begin{eqnarray*}
L= \lambda \cE^{-1} +v +u_{-1} +\lambda^{-1} uv \cE , \qquad A= \lambda^{-1} uv \cE
\end{eqnarray*}
\end{itemize}
This system comes from the Volterra chain in section \ref{vol} written in the variable $w$, that is, 
\begin{eqnarray}\label{volw}
w_t=w (w_1-w_{-1}),
\end{eqnarray}
by renaming $u(n,t)=w(2n,t)$ and $v(n,t)=w(2n-1,t)$. It is related to the Toda equation (\ref{todaeq}), written in variables
$\bar{u}$ and $\bar{v}$, by the Miura
transformation \cite{suris3}
\begin{eqnarray}\label{todvol}
\bar{u}=u v\quad \mbox{and} \quad \bar{v}=u_{-1}+v.
\end{eqnarray}

In fact, the (master) symmetries, conservation laws and local
Hamiltonian structures of this system can be easily obtained from the Volterra chain in the same way. For instance,
we can derive the first Hamiltonian operator $\cH_1$ as follows:

A symmetry flow of the Volterra chain (\ref{volw}) is
$$
w_{\tau}=w (\cS-\cS^{-1}) w Q[n]=w w_1 Q[n+1]-w w_{-1} Q[n-1], 
$$
where $Q[n]$ is the variational derivative of a conserved density for (\ref{volw}). We now write down both even and 
odd chains and rename them for the variables $u$ and $v$ accordingly. We have
\begin{eqnarray*}
 u_{\tau}=u v_1 Q[2n+1]-u v Q[2n-1] \quad \mbox{and} \quad v_{\tau}=v u Q[2n]-v u_{-1} Q[2n-2] ,
\end{eqnarray*}
that is,
\begin{eqnarray*}
 \left(\begin{array}{l} u_{\tau}\\ v_{\tau} \end{array}\right)=
 \left(\begin{array}{cc}
 0&u(\cE-1)v  \\
  v(1-\cE^{-1})u&0 \end{array}\right)
 \left(\begin{array}{l} Q[2n]\\ Q[2n-1] \end{array}\right).
\end{eqnarray*}
Using the same method, we can derive $\cH_2$ in the list. 

Such construction is valid for all scalar equations. It is tricky if the operator is nonlocal.

\subsection{The Relativistic Volterra Lattice}\label{rvol}
\begin{itemize}
\item Equation \cite{suris2, suris3, smx}:
\begin{eqnarray*}
\left\{ {\begin{array}{l}
  u_{t}= u(v-v_{-1}+uv-u_{-1}v_{-1})\\ v_{t}= v(u_1-u+u_1v_1-uv)
 \end{array} } \right.
 \end{eqnarray*}
\item Hamiltonian structure \cite{suris2}:
\begin{eqnarray*}
&&\cH_1=\left( {\begin{array}{cc}
 0&u(1-\cE^{-1})v  \\
  v(\cE-1)u&0
 \end{array} } \right),\qquad
f_1=u+v+uv\\&&
\cH_2=\left(\!\!\! {\begin{array}{cc}
 uv(1+u) \cE u-u\cE^{-1} uv(1+u)& uv(u+v+uv)\!-\!u (\cE^{-1} uv \cE^{-1}\!+\!u \cE^{-1}\!+\!\cE^{-1}v) v\\
 v(\cE uv \cE+v \cE+\cE u)u\!-\!uv(u+v+uv) & v\cE uv(1+v) -uv(1+v) \cE^{-1} v
 \end{array} } \!\!\! \right),\\
 &&\qquad  f_2=\ln u \ \mbox{or} \ f_2=\ln v
\end{eqnarray*}
\item Recursion operator:
\begin{eqnarray*}
&& \cR=\cH_{2}\cH_{1}^{-1}=\left(\begin{array}{cc}  u v_{-1} \cE^{-1}+u+v+uv & u(1+u_{-1}) \cE^{-1}+u (1+u)  \\
v(1+v_1) \cE+\frac{u_1 v(1+v_1)}{u}& u_1 v \cE +u_1 +v +u_1 v_1\end{array}\right)\\
&&\qquad +\left(\begin{array}{c}  u(v-v_{-1}+uv-u_{-1}v_{-1}) \\v(u_1-u+u_1v_1-uv) \end{array}\right) (\cS-1)^{-1}
 \left(\begin{array}{cc} \frac{1}{u} & \frac{1}{v} \end{array}\right)
\end{eqnarray*}
\item Non-trivial symmetry:
\begin{eqnarray*}
\cR\left(\begin{array}{c} u_t \\v_t \end{array}\right)=\left( \begin{array}{c}
u v (1+u) (u+u_1+u_1 v_1)+u v^2 (1+u)^2-u v_{-1}^2 (1+u_{-1})^2-u^2 v_{-1}\\ \qquad-u_{-1} u v_{-1} ( 1+u+ v_{-2}+u_{-1}+u_{-2}v_{-2});\\
u_1 v v_1 (1+2 u_1 +u_2  v_2+u_2 )+u_1^2 v+u_1 v v_1^2 (1+u_1)+v^2 u_1 (1+v_1)\\
\qquad-u v (1+v) (v+v_{-1}+u_{-1} v_{-1})-u^2 v (1+v)^2 
 \end{array}  \right)
\end{eqnarray*}
\item Lax representation 
\begin{eqnarray*}
&&M= \left( \begin{array}{cc} \lambda^2+2 u_1+2 v+1& -\lambda (2 v+1)-\lambda^{-1} (2 u_1+1) \\ -\lambda (2 u_1+1)-\lambda^{-1} (2v +1)
& \lambda^{-2} +2u_1 +2v +1\end{array}\right); \\
&&U= \left( \begin{array}{cc} -\frac{\lambda^2-\lambda^{-2}}{8}+u v +\frac{u}{2}+\frac{v}{2}&  \lambda \frac{(2v+1)}{4}+\lambda^{-1} \frac{(2u+1)}{4} \\
\lambda \frac{(2u+1)}{4}+\lambda^{-1} \frac{(2v+1)}{4} & \frac{\lambda^2-\lambda^{-2}}{8}+u v +\frac{u}{2}+\frac{v}{2}
\end{array}\right)
\end{eqnarray*}

In \cite{suris2}, the Lax representation is given in the following form:
\begin{eqnarray*}
U_t = U\ C - A\ U; \qquad W_t = W\ B - C\ W; \qquad V_t = V\ B - A\ V,
\end{eqnarray*}
where the difference operators $U, V, W, A, B$ and $C$ are given be
\begin{eqnarray*}
\begin{array}{lll} U=u+\lambda \cE^{-1}; & W=1+\lambda^{-1} v \cE; & V=1-\lambda^{-1} uv \cE \\
A=u+u_{-1} v_{-1}+v_{-1}+\lambda \cE^{-1}; & B=u+uv+v_{-1}+\lambda \cE^{-1}; & C=u+uv+v+\lambda \cE^{-1}\end{array}
\end{eqnarray*}
\end{itemize}
There exists another weakly nonlocal recursion operator
The recursion operator $\cR$ has a weakly nonlocal inverse: 
\begin{eqnarray*}
 &&\cR'=\left( \begin{array}{cc} \frac{u v}{(u_1+ v+1)^2} \cE+\frac{uv(u+1)+((1+v_{-1})^2+u)(u_1+1)}{(u+v_{-1}+1)^2 (u_1+v+1)}
 & -\frac{u(u+1)}{(u+v_{-1}+1)^2}\cS^{-1}-\frac{u(u_1+1)}{(u_1+v+1)^2}  \\
 -\frac{v (1+v)}{(u_1+ v+1)^2}\cS-\frac{v(u^2+u^2v+2 u_1 v_{-1}+u_1+u_1 v_{-1}^2)}{u (u+v_{-1}+1)(u_1+v+1)} & 
 \frac{u v}{(u+v_{-1}+1)^2}\cS^{-1}+\frac{(1+u_1)(1+v)}{(u_1+v+1)^2} \end{array}  \right)\\
&&\qquad  +\left(\begin{array}{c} -\frac{u v_{-1}}{u+v_{-1}+1}+\frac{uv}{u_1+v+1} \\ -\frac{u v}{u+v_{-1}+1}+\frac{u_1v}{u_1+v+1} \end{array}\right) (\cS-1)^{-1}
 \left(\begin{array}{cc} \frac{2}{u+v_{-1}+1} -\frac{1}{u} & \frac{2}{u_{1}+v+1}-\frac{1}{v} \end{array}\right)
\end{eqnarray*}
In fact, it commutes with $\cR$ and satisfies $\cR' (\cR+{\rm id})={\rm id}$.

It is related to the Relativistic Toda equation (\ref{rtoda}), written in variables
$\bar{u}$ and $\bar{v}$, by the Miura
transformation $\bar{u}=- u v$ and $\bar{v}=- (u+v_{-1}+1)$ \cite{suris3}. This transformation is similar to
(\ref{todvol}), which explains the name of this equation.

\subsection{The Merola-Ragnisco-Tu Lattice}\label{merola}
\begin{itemize}
\item Equation \cite{mr93c:58096, mrt94}:
\begin{eqnarray*}
\left\{ {\begin{array}{l} u_t= u_{1}-u^2v\\ v_t=  -v_{-1}+v^2u \end{array} } \right.
 \end{eqnarray*}
\item Hamiltonian structure \cite{mr93c:58096}:
\begin{eqnarray*}
 \cH=\left(\begin{array}{cc} 0 &1\\-1&0\end{array}\right), \qquad f=u_1 v-\frac{u^2v^2}{2}
\end{eqnarray*}
\item Recursion operator \cite{mr93c:58096}:
\begin{eqnarray*}
 \cR=\left(\begin{array}{cc} \cS-2 u v & -u^2\\v^2 & \cS^{-1} \end{array}\right)+2 \left(\begin{array}{c} -u \\v \end{array}\right) (\cS-1)^{-1}
 \left(\begin{array}{cc} v & u \end{array}\right)
\end{eqnarray*}
\item Non-trivial symmetry \cite{mr93c:58096}:
\begin{eqnarray*}
\cR\left(\begin{array}{c} u_t \\v_t \end{array}\right)=\left( \begin{array}{c}
 u_2-u_{1}^{2}v_{1}-u^2v_{-1}-2uvu_1+u^3v^2 \\
 -v_{-2}+v_{-1}^{2}u_{-1}+v^2u_{1}+2uvv_{-1}-u^2 v^3
 \end{array}  \right)
\end{eqnarray*}
\item Master symmetry \cite{mr93c:58096}:
\begin{eqnarray*}
\cR\left(\begin{array}{c} (n+1)u \\-n v \end{array}\right)=\left( {\begin{array}{c}
 n u_t+2u_1-2u^2 v-2 u(\cS-1)^{-1}uv  \\
 n v_t +v_{-1}+u v^2+2v(\cS-1)^{-1}uv
 \end{array} } \right)
 \end{eqnarray*}
 \item Lax representation:
\begin{eqnarray*}
M= \left( \begin{array}{cc} -1 & v \\ u & -2\lambda-u v \end{array}\right); \qquad
U= \left( \begin{array}{cc}-\lambda & -v_{-1} \\ -u & \lambda \end{array}\right) 
\end{eqnarray*}
\end{itemize}
The recursion operator $\cR$ has a weakly nonlocal inverse: 
\begin{eqnarray*}
 &&\cR^{-1}=\left( \begin{array}{cc} \frac{1}{(u_{-1} v+1)^2} \cE^{-1}  & \frac{u_{-1}^2}{(u_{-1}v+1)^2}  \\
 -\frac{v_1^2}{(u v_1+1)^2} & \frac{1}{(u v_1+1)^2}\cS-\frac{2 u_{-1}v_1}{(u_{-1}v+1) (u v_1+1)} \end{array}  \right)\\
&&\qquad  +2 \left(\begin{array}{c} \frac{u_{-1}}{u_{-1}v+1} \\ -\frac{v_{1}}{u v_1+1} \end{array}\right) (\cS-1)^{-1}
 \left(\begin{array}{cc} \frac{v_1}{uv_1+1} & \frac{u_{-1}}{u_{-1} v+1} \end{array}\right)
\end{eqnarray*}
The symmetry $(u, \ -v)^{\mbox{tr}} $ is a seed for both $\cR$ and $\cR^{-1}$.

Under the invertible transformation $t\mapsto -t$, $u\mapsto -u$ and $v \mapsto v_{1}$, this lattice transforms into
\begin{eqnarray*}
\left\{ {\begin{array}{l} u_t= -u_{1}-u^2v_1\\ v_t=  v_{-1}+v^2u_{-1} \end{array} } \right.,
 \end{eqnarray*}
which is related to the nonlinear Schr{\"o}dinger system  $u_t=u_{xx}+2 u^2 v$, $-v_t=v_{xx}+2 v^2 u$ appeared
in \cite{mr95c:58156}.

\subsection{The Kaup Lattice}
\begin{itemize}
\item Equation \cite{mr95c:58156}:
\begin{eqnarray*}
\left\{ {\begin{array}{l} u_t= (u+v) (u_1-u)\\ v_t= (u+v) (v-v_{-1}) \end{array} } \right.
 \end{eqnarray*}
\item Hamiltonian structure \cite{mr95c:58156}:
\begin{eqnarray*}
 \cH=\left(\begin{array}{cc} 0 &u+v\\-(u+v)&0\end{array}\right), \qquad f=u_1v-uv
\end{eqnarray*}
\item Recursion operator:
\begin{eqnarray*}
 &&\cR=\left(\begin{array}{cc} (u+v)\cS+u_1-u & 0 \\ 0 & (u+v)\cS^{-1}+u_1-u \end{array}\right)
 + \left(\begin{array}{c} u_t \\v_t \end{array}\right) (\cS-1)^{-1}\left(\begin{array}{cc} \frac{1}{u+v} & \frac{1}{u+v} \end{array}\right)
\\
&&\qquad  + \left(\begin{array}{c} 1 \\-1 \end{array}\right) \cS (\cS-1)^{-1}\left(\begin{array}{cc} v_{-1}-v & u_1-u \end{array}\right)
\end{eqnarray*}
\item Non-trivial symmetry:
\begin{eqnarray*}
\cR\left(\begin{array}{c} u_t \\v_t \end{array}\right)=\left( \begin{array}{c}
(u+v) (u u_1+u v_{-1}+u_1 v_1-u_1 u_2-u_2 v_1-u_1 v_{-1}) \\
(u+v) (u_{-1} v_{-2}+v_{-2} v_{-1}- u_1 v-u_{-1} v_{-1}+u_1 v_{-1}-v_{-1} v)
 \end{array}  \right)
\end{eqnarray*}
 \item Lax representation \cite{mr95c:58156}:
\begin{eqnarray*}
M= \left( \begin{array}{cc} u-\lambda & uv+\lambda (u+v)+\lambda^2 \\ 1 & v-\lambda \end{array}\right); \qquad 
U= \left( \begin{array}{cc}u & (u+\lambda)(v_{-1}+\lambda) \\ 1 & v_{-1} \end{array}\right)
\end{eqnarray*}
\end{itemize}
There exists another weakly nonlocal recursion operator
\begin{eqnarray*}
&&\cR'=\!\left(\!\!\!\! {\begin{array}{cc} \frac{u+v}{(u_{-1}+v)^2}\cS^{-1} 
& -\frac{ (u -u_{-1})}{(u_{-1}+v)^2}\\
-\frac{(v_1 -v)}{ (u+v_1)^2} &
\frac{u+v}{(u+v_1)^2}\cS+\frac{ u -u_{-1}-v_1+ v}{(u_{-1}+v) (u+v_1)} 
\end{array} } \!\!\!\!\right)
-\left(\!\!\!\begin{array}{c}1 \\ -1\end{array}\!\!\!\right) (\cS-1)^{-1}
 \left(\!\!\!\begin{array}{cc}  \frac{1}{u+v_1}-\frac{1}{u+v}
 & \frac{1}{u_{-1}+v}-\frac{1}{u+v}\end{array}\!\!\!\right)
\\&&\qquad
 -\left(\!\!\!\begin{array}{c} \frac{u-u_{-1}}{u_{-1}+v}\\ \frac{v_1-v}{u+v_1}\end{array}\!\!\!\right)
 (\cS-1)^{-1} \left(\!\!\!\begin{array}{cc} \frac{2 }{u+v_1}-\frac{1}{u+v}
 & \frac{2}{u_{-1}+v}-\frac{1}{u+v}\end{array}\!\!\!\right).
\end{eqnarray*}
The symmetry $(1, \ -1)^{\mbox{tr}} $ is a seed for both $\cR$ and $\cR'$. In fact, operator $\cR'$ is the inverse operator of $\cR$.

\subsection{The Ablowitz-Ladik Lattice}\label{AL}
  \begin{itemize}
 \item Equation \cite{ablowitz2}:
\begin{eqnarray*}
\left\{ {\begin{array}{l} u_t=(1-u v) (\alpha u_1-\beta  u_{-1}) \\ v_t= (1- uv) (\beta v_1-\alpha v_{-1})  \end{array} } \right. :=\alpha K_1 +\beta K_{-1}
 \end{eqnarray*}
\item Hamiltonian structure \cite{mr93c:58096}:
\begin{eqnarray*}
&&\cH=\left( \begin{array}{cc} 0& 1-uv \\ -(1-uv)&0\end{array}  \right),\qquad f=(\alpha u_1-\beta u_{-1}) v
\end{eqnarray*}
\item Recursion operator \cite{mr93c:58096, hssw05, hereman2}:
\begin{eqnarray*}
&& \cR=\left(\begin{array}{ll} (1-uv)~\cE - u_1 v -u v_{-1} &-uu_{1}  \\ vv_{-1}&(1-uv)~\cE^{-1}\end{array} \right)+\left(\begin{array}{c} -u\\v \end{array}\right) (\cE-1)^{-1}
\left(\begin{array}{cc} v_{-1} & u_{1} \end{array}\right) 
\\&&\qquad -\left(\begin{array}{c} (1-uv) u_{1}\\-(1-uv) v_{-1} \end{array}\right) (\cE-1)^{-1}
\left(\begin{array}{cc} \frac{v}{1-uv} & \frac{u}{1-uv} \end{array}\right)
\end{eqnarray*}
\item Non-trivial symmetry \cite{mr93c:58096}:
\begin{eqnarray*}
\cR\left(\begin{array}{c} (1-u v)  u_1 \\ -(1-u v) v_{-1} \end{array}\right)=\left( \begin{array}{l}
(1- u v) \left( (1- u_1 v_1) u_2-v u_1^2-u u_1 v_{-1} \right)\\
(1- u v) \left( -(1- u_{-1} v_{-1}) v_{-2}+ u v_{-1}^2+ u_1 v_{-1} v \right)
 \end{array}  \right)
\end{eqnarray*}
\item Master symmetry \cite{mr93c:58096}:
\begin{eqnarray*}
\cR\left(\begin{array}{c} n u \\- n v \end{array}\right)=\left( {\begin{array}{c}
 (n+1) (1- u v) u_1-u^2 v_{-1}- u(\cS-1)^{-1}u v _{-1} \\
 (1-n) (1- u v) v_{-1} + uv v_{-1}+ v (\cS-1)^{-1}u v_{-1}
 \end{array} } \right)
 \end{eqnarray*}
 \item Lax representation:
\begin{eqnarray*}
M= \left( \begin{array}{cc}\lambda &  u \\ v & \frac{1}{\lambda}\end{array}\right); \qquad 
U= \alpha \left( \begin{array}{cc} \lambda^2-u v_{-1}&  \lambda u \\ \lambda v_{-1} & 0 \end{array}\right)
+\beta \left( \begin{array}{cc} 0 & \lambda^{-1} u_{-1}\\ \lambda^{-1} v &\lambda^{-2}-u_{-1} v \end{array}\right)
\end{eqnarray*}
\end{itemize}

The coefficients for $\alpha$ and $\beta$, namely, $K_1$ and $K_{-1}$, are commuting symmetries for the equation.
The inverse of this recursion operator $\cR$ is of weakly nonlocal form:
\begin{eqnarray*}
&& \cR^{-1} =\left(\begin{array}{ll} (1-uv)~\cE^{-1}&uu_{-1}  \\ -vv_{1}& (1-uv)~\cE -uv_1-u_{-1}v \end{array} \right)
+\left(\begin{array}{c} u\\-v \end{array}\right) (\cE-1)^{-1}
\left(\begin{array}{cc} v_1 & u_{-1} \end{array}\right)\\
&&\qquad +\left(\begin{array}{c} (1-uv) u_{-1}\\-(1-uv) v_1 \end{array}\right) (\cE-1)^{-1}
\left(\begin{array}{cc} \frac{v}{1-uv} & \frac{u}{1-uv} \end{array}\right)
\end{eqnarray*} 
Both $\cR$ and $\cR^{-1}$ share a common the seed $\sigma=\left( \begin{array}{c}u\\-v \end{array}\right)$.
Starting form it,
we can generate the commuting symmetries $\cR^{-i} (\sigma)$ and $\cR^i (\sigma)$ for $i\in \mathbb{N}$.
For more Lie algebra structure among symmetries and master symmetries, we refer the reader to \cite{mr93c:58096}.

\subsection{The Bruschi-Ragnisco Lattice}
\begin{itemize}
 \item Equation \cite{bruschi1,suris2, mr90g:58048}:
\begin{eqnarray*}
\left\{ {\begin{array}{l}u_t=u_{1}v-uv_{-1}\\ v_t=v(v-v_{-1}) \end{array} } \right.
 \end{eqnarray*}
\item Hamiltonian structure \cite{bruschi1, mr93c:58096}:
\begin{eqnarray*}
&&\cH_1=\left( \begin{array}{cc}
 0&(1-\cE^{-1})v  \\ v(\cE-1)&0
 \end{array} \right),\qquad f_1=u_1 v\\
&&\cH_2=\left( \begin{array}{cc} v \cS u-u\cS^{-1} v & v(\cS-1) v\\ v (1-\cS^{-1}) v & 0
 \end{array} \right),\qquad f_2=u
\end{eqnarray*}
\item Recursion operator \cite{mr90g:58048, mr93c:58096}:
\begin{eqnarray*}
 &&\cR=\cH_2 \cH_1^{-1}=\left( \begin{array}{cc}
 v\cE&u_1+u\cE^{-1}+(u_1v-uv_{-1})(\cE-1)^{-1}\frac{1}{v}  \\
 0&v\cE^{-1}+v(v-v_{-1})(\cE-1)^{-1}\frac{1}{v} \end{array}  \right)\\
&&\quad = \left( \begin{array}{cc}
 v\cE&u_1+u\cE^{-1}\\
 0&v\cE^{-1}\end{array}  \right) +\left(\begin{array}{c} u_t \\v_t \end{array}\right) (\cS-1)^{-1}
 \left(\begin{array}{cc} 0 & \frac{1}{v} \end{array}\right)
\end{eqnarray*}
\item Non-trivial symmetry \cite{mr93c:58096}:
\begin{eqnarray*}
\cR\left(\begin{array}{c} u_t \\v_t \end{array}\right)=\left( \begin{array}{c}
 vv_{1}u_{2}-v_{-2}v_{-1}u\\
v(vv_{-1}-v_{-1}v_{-2})
 \end{array}  \right)
\end{eqnarray*}
\item Master symmetry \cite{mr93c:58096}:
\begin{eqnarray*}
\cR\left(\begin{array}{c} u \\ v \end{array}\right)=\left( {\begin{array}{c}
 n u_t+ 2 u_1 v+u_{-1} u  \\
 n v_t +v v_{-1}
 \end{array} } \right)
 \end{eqnarray*}
\end{itemize}

The recursion operator $\cR$ has a weakly nonlocal inverse: 
\begin{eqnarray*}
 &&\cR^{-1}=\cH_1 \cH_2^{-1}=\left( \begin{array}{cc}
  \cE^{-1} \frac{1}{v} &-\cE^{-1} \frac{u}{v^2} -\frac{u}{v^2} +(\frac{u_{-1}}{v_{-1}} -\frac{u}{v}) (\cE-1)^{-1}\frac{1}{v}  \\
 0& v \cE \frac{1}{v^2} -\frac{1}{v} +\frac{1}{v_1}+( \frac{v}{v_{1}}-1) (\cE-1)^{-1}\frac{1}{v} \end{array}  \right)\\
&&\qquad = \left( \begin{array}{cc}
\cE^{-1} \frac{1}{v} &-\cE^{-1} \frac{u}{v^2} -\frac{u}{v^2}\\
  0& v \cE \frac{1}{v^2} -\frac{1}{v} +\frac{1}{v_1}\end{array}  \right) +\left(\begin{array}{c} \frac{u_{-1}}{v_{-1}} -\frac{u}{v} \\\frac{v}{v_{1}}-1 \end{array}\right) (\cS-1)^{-1}
 \left(\begin{array}{cc} 0 & \frac{1}{v} \end{array}\right)
\end{eqnarray*}
However, recursion operators $\cR$ and $\cR^{-1}$ have different seeds similar to the Relativistic Toda system in section \ref{secrtoda}. 
The starting point for operator $\cR$ is the equation itself while
the seed for $\cR^{-1}$ is $\sigma=\left( \begin{array}{l}\frac{u_{-1}}{v_{-1}} -\frac{u}{v} 
\\ \frac{v}{v_{1}}-1 \end{array}\right)$. Operator $\cR^{-1}$ acting on the equation itself and $\cR$ on $\sigma$ do not generate new symmetries.
We refer the reader to \cite{mr93c:58096} for more Lie algebra structure among symmetries and master symmetries.

In fact, the Bruschi-Ragnisco lattice is trivially solvable, so are its symmetry flows. 
The equation for the second component $v_t$ is independent of the first component $u$. Besides,
the scalar lattice $v_t=v (v-v_{-1})$ and $v_t=\frac{v}{v_{1}}-1$ can be linearised into $w_t=w_{-1}$ and $w_t=w_1$, respectively
by the transformations $v=-\frac{w_{-1}}{w}$ and $v=-\frac{w}{w_1}$, respectively. Once it is solved, the equation for $u$ 
is then linear. 


\subsection{The Kaup-Newell lattice}\label{seckn}
\begin{itemize}
\item Equation \cite{tsuchida1}:
\begin{eqnarray*}
\left\{\begin{array}{l}
 u_t=a\left( \frac{u_1}{1-u_{1}v_{1}}-\frac{u}{1-uv}\right)+b\left(\frac{u}{1+uv_{1}}-\frac{u_{-1}}{1+u_{-1}v}\right)\\
 v_t=a\left( \frac{v}{1-u v}-\frac{v_{-1}}{1-u_{-1} v_{-1}}\right)+b\left(\frac{v_1}{1+uv_{1}}-\frac{v}{1+u_{-1}v}\right)
\end{array}  \right. :=a K_1 +b K_{-1}
\end{eqnarray*}
\item Hamiltonian structure \cite{tsuchida1}:
\begin{eqnarray*}
\cH=\left( {\begin{array}{cc}
 0&\cE-1  \\1-\cE^{-1}&0\end{array} } \right), \qquad  f=-a\ln(1-uv)+b\ln(1+uv_{1})
\end{eqnarray*}
\item Recursion operator:
\begin{eqnarray*}
&&\cR=\cH \J\!=\!\left(\!\!\!\! {\begin{array}{cc}
 -\frac{1}{(1-u_1v_1)^2} \cS+\frac{1}{(1-uv)^2}-\frac{2 u_1 v}{(1-u_1v_1)(1-uv)}
&-\frac{u_1^2}{(1-u_1 v_1)^2} \cS+\frac{u^2}{(1-u v)^2}-\frac{2 u u_1}{(1-uv)(1-u_1 v_1)} \\
-\frac{v_{-1}^2}{(1-u_{-1} v_{-1})^2} \cS^{-1}-\frac{v^2}{(1-u v)^2}&
-\frac{1}{(1-u_{-1}v_{-1})^2} \cS^{-1}+\frac{1-2uv}{(1-uv)^2}\end{array} } \!\!\!\!\right)
\\&&\qquad
 -2 K_1 (\cS-1)^{-1} 
\left(\begin{array}{cc} \frac{v}{1-u v} & \frac{u}{1-u v}\end{array}\right),
\end{eqnarray*}
where the symplectic operator $\J$ is defined by
\begin{eqnarray*}
 \J=\left( {\begin{array}{cc}
 0&\frac{1}{1-u v} \\-\frac{1}{1-u v}&0\end{array} } \right)
 -\left(\begin{array}{c} \frac{v}{1-u v}\\ \frac{u}{1-u v}\end{array}\right) (\cS+1) (\cS-1)^{-1} 
 \left(\begin{array}{cc} \frac{v}{1-u v} & \frac{u}{1-u v}\end{array}\right)
\end{eqnarray*}
\item Non-trivial symmetry :
\begin{eqnarray*}
\cR\left(K_1\right)=\left( \begin{array}{c} \frac{1}{(1-u_1 v_1)^2}(u_1-\frac{u_2 }{1-u_2 v_2}-\frac{u_1^2 v}{1-u v})
-\frac{1}{(1-u v)^2}(u-\frac{u_1 }{1-u_1 v_1}-\frac{u^2 v_{-1}}{1-u_{-1} v_{-1}})\\
\frac{1}{(1-u v)^2}(v-\frac{u_1 v^2}{1-u_1 v_1}-\frac{v_{-1}}{1-u_{-1} v_{-1}})
-\frac{1}{(1-u_{-1} v_{-1})^2}(v_{-1}-\frac{u v_{-1}^2 }{1-u v}-\frac{v_{-2}}{1-u_{-2} v_{-2}})
 \end{array}  \right)
\end{eqnarray*}
\item Lax representation \cite{tsuchida1}:
\begin{eqnarray*}
&&M= \left( \begin{array}{cc} \lambda+(1-\lambda) uv&  u \\ (1-\lambda) v & 1 \end{array}\right); \\
&&U= a\left( \begin{array}{cc} \frac{\lambda-1}{2} &  \frac{u}{1-uv} \\  \frac{(1-\lambda) v_{-1}}{1-u_{-1} v_{-1}} & - \frac{\lambda-1}{2}\end{array}\right) 
+\frac{b}{2 \lambda (1+u_{-1}v)} \left( \begin{array}{cc}(\lambda-1)(1-u_{-1}v) &  2 u_{-1} \\  2 (1-\lambda) v & (1-\lambda)(1-u_{-1}v)\end{array}\right) 
\end{eqnarray*}
\end{itemize}

The recursion operator $\cR$ has a seed $\sigma=\left(\begin{array}{c} -u\\v \end{array}\right)$ and $\cR (\sigma)=K_1$ taking integration constant being zero. 
Similar to the Ablowitz-Ladik Lattice in section \ref{AL}, the coefficients for $a$ and $b$, namely, $K_1$ and $K_{-1}$, are commuting symmetries for the equation.
Indeed, there exists another weakly nonlocal recursion operator
\begin{eqnarray*}
&&\cR'=\cH \J'=\!\left(\!\!\!\! {\begin{array}{cc} \frac{1}{(1+u_{-1}v)^2} \cS^{-1} -\frac{1+2 u v_1}{(1+u v_1)^2} & -\frac{u^2}{(1+u v_1)^2} \cS
+\frac{u_{-1}^2}{(1+u_{-1} v)^2}-\frac{2 u u_{-1}}{(1+u_{-1}v) (1+u v_1)}\\
-\frac{v^2}{(1+u_{-1} v)^2} \cS^{-1}-\frac{v_1^2}{(1+u v_1)^2} &
\frac{1}{(1+u v_1)^2} \cS-\frac{1}{(1+u_{-1} v)^2}-\frac{2 u_{-1} v_1}{(1+u_{-1} v) (1+u v_1)}
\end{array} } \!\!\!\!\right)
\\&&\qquad
 -2 K_{-1} (\cS-1)^{-1} 
\left(\begin{array}{cc} \frac{v_1}{1+u v_1} & \frac{u_{-1}}{1+u_{-1} v}\end{array}\right),
\end{eqnarray*}
where the symplectic operator $\J'$ is
\begin{eqnarray*}
&&\J'=\left( {\begin{array}{cc}
 0&\frac{1}{1+u v_1} (\cS-u_{-1} v_1) \frac{1}{1+u_{-1}v} \\ \frac{1}{1+u_{-1}v} (u_{-1} v_1-\cS^{-1}) \frac{1}{1+u v_1} &0\end{array} } \right)\\
&&\qquad
 -\left(\begin{array}{c} \frac{v_1}{1+u v_1}\\ \frac{u_{-1}}{1+u_{-1} v}\end{array}\right) (\cS+1) (\cS-1)^{-1} 
 \left(\begin{array}{cc} \frac{v_1}{1+u v_1} & \frac{u_{-1}}{1+u_{-1} v}\end{array}\right)
\end{eqnarray*}
Again $\sigma$ is a seed for $\cR'$ and $\cR'(\sigma)=K_{-1}$ taking
$$(\cS-1)^{-1} \left(-\frac{u v_1}{1+u v_1}+\frac{u_{-1}v}{1+u_{-1} v} \right)=-\frac{u_{-1} v}{1+u_{-1} v}. $$ 
By direct calculation, we have
$\J'\cR=\J\cR'=\J'-\J$. Thus these two recursion operators satisfy the relation
$\cR' \cR=\cR\cR'=\cR'-\cR$, that is, $(\cR'-{\rm id})(\cR-{\rm id})=(\cR-{\rm id})(\cR'-{\rm id})={\rm id}$.

\subsection{The Chen-Lee-Liu lattice}\label{seccll}
\begin{itemize}
\item Equation \cite{tsuchida1}:
\begin{eqnarray*}
\left\{\begin{array}{l}
u_t=a(1+uv)(u_1-u)+b(1+u_{-1}v)^{-1} (u-u_{-1})\\
v_t=a(1+uv)(v-v_{-1})+b(1+uv_1)^{-1}(v_1-v)
 \end{array}  \right.:=a K_1 +b K_{-1}
 \end{eqnarray*}
\item Hamiltonian structure:
\begin{eqnarray*}
&&\cH=\left( {\begin{array}{cc}
 0& 1+uv  \\-(1+uv)&0\end{array} } \right), \qquad  f=a (u v_{-1}-uv)+b\ln\frac{1+uv}{1+uv_1}
\end{eqnarray*}
\item Recursion operator:
\begin{eqnarray*}
&&\cR=\cH_2 \cH^{-1} =\left( {\begin{array}{cc}(1+uv)\cS -2uv+u_1v+u v_{-1}
 &u u_1-u^2 \\ v^2-v v_{-1} & (1+uv)\cS^{-1} \end{array} } \right)\\
&&\qquad
+K_1 (\cS-1)^{-1} \left(\begin{array}{cc} \frac{v}{1+uv} & \frac{u}{1+uv}\end{array}\right)
-\left(\begin{array}{c} u \\ -v\end{array}\right) (\cS-1)^{-1} 
\left(\begin{array}{cc} v-v_{-1} & u-u_1 \end{array}\right), 
\end{eqnarray*}
where Hamiltonian operator $\cH_2$ is given by
\begin{eqnarray*}
&&\cH_2=\left( {\begin{array}{cc}
 0&(1+uv)\left(\cS (1+uv)-uv+u_1v\right) \\ \left(uv-u_1v -(1+uv)\cS^{-1}\right)(1+uv) &0\end{array} } \right)\\
&&\qquad
-K_1 (\cS-1)^{-1} \left(\begin{array}{cc} u & -v\end{array}\right)
-\left(\begin{array}{c} u \\ -v\end{array}\right)\cS (\cS-1)^{-1} K_1^{\mbox{tr}}
\end{eqnarray*}
\item Non-trivial symmetry :
\begin{eqnarray*}
\cR\left(K_1\right)=\left( \begin{array}{c} (1+uv)(u_1u_2v_1+u_2+u_1^2 v+u u_1 v_{-1}+u^2v-u_1-u^2 v_{-1}-2 u u_1 v -u_1^2 v_1)\\
(1+uv) (u_1 v^2+2 u v_{-1} v +u_{-1} v_{-1}^2+ v_{-1}-v_{-2}-u v^2 -u v_{-1}^2 -u_1 v_{-1} v-u_{-1} v_{-2} v_{-1})
\end{array}  \right)
\end{eqnarray*}
\item Lax representation \cite{tsuchida1}:
\begin{eqnarray*}
M= \left( \begin{array}{cc} \lambda+uv&  u \\ (1-\lambda) v & 1 \end{array}\right); \qquad
U= a \left( \begin{array}{cc}\lambda-1+u v_{-1} &  u \\ (1-\lambda) v_{-1} & 0
\end{array}\right) +\frac{b}{\lambda(1+u_{-1}v)}\left( \begin{array}{cc}u_{-1} v &  u_{-1} \\ (1-\lambda) v & 1-\lambda
\end{array}\right) 
\end{eqnarray*}
\end{itemize}
The coefficients for $a$ and $b$, namely, $K_1$ and $K_{-1}$, are commuting symmetries for the equation.
The above recursion operator $\cR$ has a seed $\sigma=\left(\begin{array}{c} -u\\v \end{array}\right)$ and $\cR (\sigma)=K_1$.

There exists another weakly nonlocal recursion operator
\begin{eqnarray*}
&&\cR'=\cH_2'\cH^{-1}=\!\left(\!\!\!\! {\begin{array}{cc} \frac{1+uv}{(1+u_{-1}v)^2}\cS^{-1} 
& -\frac{u_{-1} (u -u_{-1})}{(1+u_{-1}v)^2}\\
-\frac{v_1 (v_1 -v)}{ (1+uv_1)^2} &
\frac{1+uv}{(1+uv_1)^2}\cS+\frac{v_1 u -2 u_{-1}v_1+ u_{-1} v}{(1+u_{-1}v) (1+uv_1)} 
\end{array} } \!\!\!\!\right)
\\&&\qquad
 -K_{-1} (\cS-1)^{-1} \left(\!\!\!\begin{array}{cc} \frac{2 v_1}{1+uv_1}-\frac{v}{1+uv}
 & \frac{2 u_{-1}}{1+u_{-1}v}-\frac{u}{1+uv}\end{array}\!\!\!\right)\\
&&\qquad
-\left(\!\!\!\begin{array}{c}-u \\ v\end{array}\!\!\!\right) (\cS-1)^{-1}
 \left(\!\!\!\begin{array}{cc}  \frac{v_1}{1+uv_1}-\frac{v}{1+uv}
 & \frac{u_{-1}}{1+u_{-1}v}-\frac{u}{1+uv}\end{array}\!\!\!\right),
\end{eqnarray*}
where the Hamiltonian operator $\cH_2'$ is
\begin{eqnarray*}
&&\cH'_2=\left( {\begin{array}{cc}
 0&\frac{1+uv}{1+u_{-1}v}\left(\cS^{-1} \frac{1+uv}{1+uv_1}+\frac{v_1 (u -u_{-1})}{1+uv_1}\right) \\ 
 -\left(\frac{1+uv}{1+uv_1}\cS+\frac{v_1 (u -u_{-1})}{1+uv_1}\right)\frac{1+uv}{1+u_{-1}v}
 &0\end{array} } \right)\\
&&\qquad
-K_{-1} (\cS+1) (\cS-1)^{-1} K_{-1}^{\mbox{tr}} -K_{-1} \cS (\cS-1)^{-1} \left(\begin{array}{cc} -u & v\end{array}\right)
-\left(\begin{array}{c} -u \\ v\end{array}\right) (\cS-1)^{-1} K_{-1}^{\mbox{tr}}
\end{eqnarray*}
Again $\sigma$ is a seed for $\cR'$. In fact, operator $\cR'$ is the inverse operator of $\cR$.

\subsection{The Ablowitz-Ramani-Segur (Gerdjikov-Ivanov) lattice}\label{secgi}
\begin{itemize}
\item Equation \cite{tsuchida1}:
\begin{eqnarray*}
\left\{\begin{array}{l}
u_t=(a u_1-b u_{-1}) (1+uv) (1-uv_1)\\
v_t= (b v_1-a v_{-1})(1+uv)(1-u_{-1}v)
 \end{array}  \right.:=a K_1+b K_{-1}
\end{eqnarray*}
\item Symplectic operator:
\begin{eqnarray*}
&&\J=\left( {\begin{array}{cc}
0&\frac{1}{1-u v_1}\cS - \frac{1}{1+u v} \\ \frac{1}{1+uv} -\cS^{-1} \frac{1}{1-u v_1} &0\end{array} } \right) 
\end{eqnarray*}
and we have
\begin{eqnarray*}
 \J (aK_1+b K_{-1})=\delta_{(u,v)} \left(a (u v_{-1}-uv-uvu_1v_1)+b (u_{-1}v_1-u v_1 +u_{-1}uvv_1) \right) 
\end{eqnarray*}
\item Hamiltonian structure:
\begin{eqnarray*}
&&\cH=\left( {\begin{array}{cc}
 0& (1+uv)(\cS (1+uv)+uv_{-1} )(1-u_{-1}v) \\ -(1-u_{-1}v) ((1+uv)\cS^{-1}+uv_{-1})(1+uv) &0\end{array} } \right)\\
&&\qquad
-K_1 \cS (\cS-1)^{-1} 
 \left(\begin{array}{cc}u & -v \end{array}\right)-\left(\begin{array}{c}u \\ -v \end{array}\right)(\cS-1)^{-1} K_1^{\mbox{tr}}
\end{eqnarray*}
\item Recursion operator:
\begin{eqnarray*}
&&\cR=\left( {\begin{array}{cc}
\begin{array}{c}(1+uv)(1-uv_1)\cS+u_1v-u_1v_1\\+uv_{-1}-uv(1+u_{-1}v_{-1}+2u_1v_1)\end{array}&\begin{array}{c} -u u_1(1+uv)\cS-u^2(1+u_{-1}v_{-1})\\ 
+\frac{1-u v_1}{1-u_{-1}v} u_1 (u-u_{-1}-2 u u_{-1}v)\end{array}\\
\\-(1-u_{-1}v)v_{-1}v-(1+uv)v_{-1}v \cS^{-1} &
(1+uv)(1-u_{-1}v)\cS^{-1}+uv u_{-1}v_{-1}\end{array} } \right)\\ \\
&&\qquad
+\left(\begin{array}{c} u_1 (1+uv)(1 -u v_1)\\ -v_{-1}(1+uv)(1-u_{-1}v)\end{array}\right)  (\cS-1)^{-1} 
\left(\begin{array}{cc} \frac{v}{1+u v}-\frac{v_1}{1-u v_1} & \frac{u}{1+uv}-\frac{u_{-1}}{1-u_{-1} v}\end{array}\right)\\
&&\qquad- \left(\begin{array}{c} u \\ -v\end{array}\right)  (\cS-1)^{-1} 
\left(\begin{array}{cc}v -v_{-1}+u_{-1} v_{-1}v +u_1 v v_1 & u -u_1+u u_{-1} v_{-1}+u u_1 v_1\end{array}\right)\\
&&\qquad=\cH \J +\left(\begin{array}{cc}1&0\\0&1\end{array}\right)
\end{eqnarray*}
\item Non-trivial symmetry $\cR(K_1)$:
\begin{eqnarray*}
\left(\!\!\! \begin{array}{c} (1+uv)(1-u v_1)\left((1+u_1 v_1) (1-u_1 v_2) u_2-u_1^2 v_1 (1+uv) +u u_1 v_{-1} (1-u_{-1}v)-u_1 v(u-u_1)\right)\\
(1\!+\!uv)(u_{-1}v\!-\!1) \left(\! (1\!+\! u_{-1}v_{-1})(1\!-\!u_{-2}v_{-1}) v_{-2}\!-\!u_{-1} v_{-1}^2(1\!+\!uv)\!+\!u_1 v_{-1} v(1\!-\!u v_1)\!+\!u v_{-1} (v_{-1}\!-\!v)\! \right)
\end{array}\!\!\!\right)
\end{eqnarray*}
\item Lax representation \cite{tsuchida1}:
\begin{eqnarray*}
&&M= \left( \begin{array}{cc} \lambda+uv&  u \\ (1-\lambda) (1-uv_1) v & 1-u v_1 \end{array}\right); \\
&&U=a \left( \begin{array}{cc}\lambda+u v_{-1}(1-u_{-1}v) &  u \\ (1-\lambda) (1-u_{-1}v) v_{-1} & -uv
\end{array}\right)\\
&&\qquad+\frac{b}{\lambda} \left( \begin{array}{cc}u_{-1} v -\lambda&  u_{-1}\\ (1-\lambda)(1-u_{-1}v) v &  (1-\lambda) (1-u_{-1}v) -\lambda u_{-1}v_1(1+uv)
\end{array}\right)
\end{eqnarray*}
\end{itemize}
The equation given in \cite{tsuchida1} is 
\begin{eqnarray*}
\left(\begin{array}{l}u_t\\v_t\end{array}\right)=\left(\begin{array}{l}(b-a)u\\(a-b)v
\end{array}  \right) +a K_1+b K_{-1}.
\end{eqnarray*}
Since the vector  $\sigma=\left(\begin{array}{c} -u\\v \end{array}\right)$ commutes with both $K_1$ and $K_{-1}$, we removed
this term in our consideration.

There exists another weakly nonlocal recursion operator 
\begin{eqnarray*}
&&\cR'=\left( {\begin{array}{cc}  (1+uv)(1-u v_1)\cS^{-1}+uv u_{-1}v_{1}  & \begin{array}{c} u u_{-1} (1+uv)\cS+u u_{-2} (1+u_{-1}v_{-1})\\ 
-\frac{1-u v_1}{1-u_{-1}v} u_{-1} (u-u_{-1}-2 u u_{-1}v)\end{array} \\ \\
(1-u_{-1}v)v_{1}v+(1+uv)v_{1}v \cS^{-1} & \begin{array}{c}(1+uv)(1-u_{-1} v)\cS+u v_1-2 u_{-1} u v v_1\\-u_{-1} v_1+u_{-1} v-u_{-2} v(1+u_{-1}v_{-1})\end{array} 
\end{array} } \right)\\ \\
&&\qquad
+\left(\begin{array}{c} -u_{-1} (1+uv)(1 -u v_1)\\ v_{1}(1+uv)(1-u_{-1}v)\end{array}\right)  (\cS-1)^{-1} 
\left(\begin{array}{cc} \frac{v}{1+u v}-\frac{v_1}{1-u v_1} & \frac{u}{1+uv}-\frac{u_{-1}}{1-u_{-1} v}\end{array}\right)\\
&&\qquad+ \left(\begin{array}{c} u \\ -v\end{array}\right)  (\cS-1)^{-1} 
\left(\begin{array}{cc}v_2(1+u_1 v_1)+ v_1 (u_{-1} v -1) & u_{-2} (1 +u_{-1} v_{-1}) +u_{-1} (u v_1-1)\end{array}\right)\\
&&\qquad=\cH' \J +\left(\begin{array}{cc}1&0\\0&1\end{array}\right)
\end{eqnarray*}
where the Hamiltonian operator $\cH'$ is
\begin{eqnarray*}
&&\cH'=\left( {\begin{array}{cc} 0& -(1+uv)(1-u_{-1}v) \\ (1+uv)(1-u_{-1}v) &0\end{array} } \right)\\
&&\qquad
-K_{-1} \cS (\cS-1)^{-1} \left(\begin{array}{cc} u & -v\end{array}\right)
-\left(\begin{array}{c} u \\ -v \end{array}\right) (\cS-1)^{-1} K_{-1}^{\mbox{tr}}
\end{eqnarray*}
Operator $\cR'$ is the inverse operator of $\cR$. The vector $\sigma$ is them seed for both of them and $\cR'(K_{-1})$ is 
\begin{eqnarray*}
\left(\!\!\!\! \begin{array}{c}(1-uv_1)(1+uv)\left((1+u_{-1}v_{-1})(1-u_{-1}v)u_{-2}-uu_{-1}v_{2}(1+u_{1}v_{1})-u_{-1}^2v_{1}(1+uv)+u_{-1}(u_{-1}v+uv_{1})\right)\\
(1-u_{-1}v)(1+uv)\left(-(1+u_{1}v_{1})(1-uv_{1})v_{2}+u_{-2}vv_{1}(1+u_{-1}v_{-1})+u_{-1}v_{1}(v_{1}vu+v_{1}-v)-uv_{1}^2\right)
\end{array}\!\!\!\!\right)
\end{eqnarray*}

\subsection{The Heisenberg ferromagnet lattice}\label{heisen}
\begin{itemize}
 \item Equation \cite{sklyanin}:
\begin{eqnarray*}
\left\{ {\begin{array}{l} u_t= (u-v) (u-u_1) (u_1-v)^{-1}\\
   v_t= (u-v) (v_{-1}-v) (u-v_{-1})^{-1}  \end{array} } \right.
 \end{eqnarray*}
\item Hamiltonian structure:  
\begin{eqnarray*}
 &&\cH=\left(\begin{array}{cc}  0& (u-v)^2 \\
 -(u-v)^2&0\end{array}\right), \qquad f=\ln (u-v)-\ln(u-v_{-1})
\end{eqnarray*}
\item Symplectic operator:
\begin{eqnarray*}
 &&\J=\left( {\begin{array}{cc}
 0 & -(u-v_{-1})^{-2} \cS^{-1} +\frac{(u-u_1) (v-v_{-1})}{(u-v)^2 (u-v_{-1}) (u_1-v)} \\
(u_1-v)^{-2} \cS -\frac{(u-u_1) (v-v_{-1})}{(u-v)^2 (u-v_{-1}) (u_1-v)} & 0
 \end{array} } \right)\\
 &&\quad- \left(\begin{array}{c} \frac{v-v_{-1}}{(u-v) (u-v_{-1})}\\ \frac{u-u_1}{(u-v) (u_1-v)} \end{array}\right) (\cS+1) (\cS-1)^{-1}
 \left(\begin{array}{cc} \frac{v-v_{-1}}{(u-v) (u-v_{-1})} & \frac{u-u_1}{(u-v) (u_1-v)}  \end{array}\right)
\end{eqnarray*}
\item Recursion operator:
\begin{eqnarray*}
&& \cR
 =\cH \J=\left( {\begin{array}{cc}
 \frac{(u-v)^2}{(u_1-v)^{2}} \cS -\frac{2 (u-u_1) (v-v_{-1})}{(u-v_{-1}) (u_1-v)}& -\frac{(u-u_1)^2}{ (u_1-v)^{2}}  \\
 \frac{(v-v_{-1})^2}{ (u-v_{-1})^{2}} & \frac{(u-v)^2}{ (u-v_{-1})^{2}} \cS^{-1}
 \end{array} } \right)\\
 &&\quad-2 \left(\begin{array}{c}u_t  \\ v_t \end{array}\right) (\cS-1)^{-1}
 \left(\begin{array}{cc} \frac{v-v_{-1}}{(u-v) (u-v_{-1})} & \frac{u-u_1}{(u-v) (u_1-v)}  \end{array}\right)
\end{eqnarray*} 
\item Non-trivial symmetry:
\begin{eqnarray*}
\cR\left(\begin{array}{c} u_t \\v_t \end{array}\right)=\left( \begin{array}{c}
\frac{(u-v)}{(u_1-v)^2} \left( \frac{(u-v) (u_1-v_1) (u_1-u_2)}{(u_2-v_1)}+\frac{(u-u_1)^2 (v_{-1}-v)}{(u-v_{-1})}\right)
 \\
\frac{(u-v)}{(u-v_{-1})^2} \left( \frac{(u-v) (u_{-1}-v_{-1}) (v_{-2}-v_{-1})}{(u_{-1}-v_{-2})}+\frac{(v-v_{-1})^2 (u-u_1)}{(u_1-v)}\right)
 \end{array}  \right)
\end{eqnarray*}
 \item Lax representation:
 \begin{eqnarray*}
 &&M= \left( \begin{array}{cc} \lambda-2 u (u-v)^{-1}& -2 (u-v)^{-1} \\ 2 uv (u-v)^{-1}
& \lambda+2 v (u-v)^{-1}\end{array}\right); \\
&&U= \lambda^{-1} (u-v_{-1})^{-1}\left( \begin{array}{cc} u+v_{-1} & 2 \\ -2 u v_{-1} & -(u+v_{-1})\end{array}\right)
\end{eqnarray*}
\end{itemize}
The recursion operator $\cR$ has a weakly nonlocal inverse:
\begin{eqnarray*}
&& \cR^{-1}
 =\cH \J'=\left( {\begin{array}{cc}
 \frac{(u-v)^2}{(u_{-1}-v)^{2}} \cS^{-1}& \frac{(u-u_{-1})^2}{ (u_{-1}-v)^{2}}  \\
 -\frac{(v-v_{1})^2}{ (u-v_{1})^{2}}  & \frac{(u-v)^2}{ (u-v_{1})^{2}} \cS -\frac{2 (u-u_{-1}) (v-v_{1})}{(u-v_{1}) (u_{-1}-v)}
 \end{array} } \right)\\
 &&\quad-2 \left(\begin{array}{c} \frac{(u-v)(u_{-1}-u)}{u_{-1}-v}  \\ \frac{(u-v) (v-v_1)}{u-v_1}  \end{array}\right) (\cS-1)^{-1}
 \left(\begin{array}{cc} \frac{v-v_{1}}{(u-v) (u-v_{1})} & \frac{u-u_{-1}}{(u-v) (u_{-1}-v)}  \end{array}\right),
\end{eqnarray*} 
where the symplectic operator $\J'$ is given by
\begin{eqnarray*}
 &&\J'=\left( {\begin{array}{cc}
 0 & -(u-v_{1})^{-2} \cS +\frac{(u-u_{-1}) (v-v_{1})}{(u-v)^2 (u-v_{1}) (u_{-1}-v)} \\
(u_{-1}-v)^{-2} \cS^{-1} -\frac{(u-u_{-1}) (v-v_{1})}{(u-v)^2 (u-v_{1}) (u_{-1}-v)} & 0
 \end{array} } \right)\\
 &&\quad+ \left(\begin{array}{c} \frac{v-v_{1}}{(u-v) (u-v_{1})}\\ \frac{u-u_{-1}}{(u-v) (u_{-1}-v)} \end{array}\right) (\cS+1) (\cS-1)^{-1}
 \left(\begin{array}{cc} \frac{v-v_{1}}{(u-v) (u-v_{1})} & \frac{u-u_{-1}}{(u-v) (u_{-1}-v)}  \end{array}\right)
\end{eqnarray*}

The Heisenberg ferromagnet lattice is a special case of the following
{\bf Landau-Lifshitz (Sklyanin) chain} \cite{sklyanin, adler_LL, mr91k:58116}:
\begin{equation}\label{YSh}
\left\{ {\begin{array}{l} u_t= a\left(\frac{2h}{u_1-v}+h_v\right)+b\left(\frac{2h}{u_{-1}-v}+h_v\right)\\
v_t= a\left(\frac{2h}{u-v_{-1}}-h_u\right)+b\left(\frac{2h}{u-v_1}-h_u\right) \end{array} } \right.
\end{equation}
 where 
 \[
  h(u,v)=\frac{i}{4}(K_1 (1-uv)^2-K_2 (1+uv)^2+K_3 (u+v)^2),\qquad K_1K_2K_3\ne 0, \ \ K_n\in\bbbr
 \]
by specifying $a=1$, $b=0$ and $h(u,v)=\frac{1}{2} (u-v)^2$.
Notice that the coefficients of $a$ and $b$ in the Landau-Lifshitz (Sklyanin) chain (\ref{YSh}) commute.

The Lax representation for equation (\ref{YSh}) is given in \cite{adler_LL} by
\begin{eqnarray*}
U=  i\sum_{k=1}^3 a N^+_k(\lambda) S_k(u,v_{-1})\sigma_k+b  N^-_k(\lambda) S_k(v,u_{-1})\sigma_k; \ 
M= \frac{1}{\sqrt{\langle \bf {SKS}\rangle}}\left(\!I- \sum_{k=1}^3  M_k(\lambda) S_k(u,v)\sigma_k\!\right) .
\end{eqnarray*}
Here $\sigma_k$ are the Pauli matrices
\[
 \sigma_1=\left(\begin{array}{rr}
                 0&1\\1&0
                \end{array}
\right),\qquad 
\sigma_2=\left(\begin{array}{rr}
                 0&-i\\i&0
                \end{array}
\right),\qquad 
\sigma_3=\left(\begin{array}{rr}
                 1&0\\0&-1
                \end{array}
\right),
\]
vector function $S(p,q)$ is defined as following
\[
  (S_1(p,q),S_2(p,q),S_3(p,q))=  \left(\frac{1-p\, q}{p-q},\frac{i+ip\, q}{p-q},\frac{p+q}{p-q}
  \right),\]
               satisfying $S_1^2+S_2^2+S_3^2=1$,
               \[
                \langle {\bf
             SKS}\rangle=\sum_{l=1}^3  K_l S_l^2(u,v)
               \]
and $M_l(\lambda),N^{\pm}_l(\lambda)$ can be expressed in terms of Jacobi elliptic functions of a spectral parameter $\lambda$:
\[
 M_1(\lambda)=\sqrt{1-K_1K_2^{-1}}\mbox{sn} (\lambda ,\kappa), \ M_2(\lambda)=\sqrt{1-K_2K_1^{-1}}\mbox{cn} (\lambda ,\kappa),
 \ M_3(\lambda)=\sqrt{1-K_3K_1^{-1}}\mbox{dn} (\lambda ,\kappa) \]
\[
 N_1^{\pm}(\lambda)=\frac{\sqrt{K_2K_3}}{2}M_1(\lambda\pm \mu),\quad 
 N_2^{\pm}(\lambda)=\frac{\sqrt{K_1K_3}}{2\kappa}M_2(\lambda\pm \mu),\quad
 N_3^{\pm}(\lambda)=\frac{\sqrt{K_1K_2}}{2}M_3(\lambda\pm \mu),\]
 where $\kappa$ and $\mu$ are defined by equations
 \[ \kappa=\sqrt{\frac{K_3 (K_1-K_2)}{K_2 (K_1-K_3)}},\quad \mbox{cn}(\mu,\kappa)=\frac{K_1}{K_3} .
\]
Instead of explicit uniformisation of the elliptic curve, one can use identities
\[
 \frac{M_l^2-1}{K_l}=\frac{M_j^2-1}{K_j},\quad N_l^{\pm}=\frac{K_1}{2(M_1^2-1)}(M_l\pm M_j M_k),\quad l,j,k\in \{1,2,3\},\ l\ne j\ne k\ne l.
\]
One local Hamiltonian operator for equation (\ref{YSh}) \cite{adler_LL} is
\begin{eqnarray*}
 \cH=h(u,v) \left(\begin{array}{cc} 0 &1\\-1&0\end{array}\right),\qquad f=a \ln \frac{h(u,v)}{(u_{1}-v)^2} +b \ln \frac{h(u,v)}{(u_{-1}-v)^2}  .
\end{eqnarray*}
Knowing the Lax representation, we can, in principle, compute its recursion operators (as we did in section \ref{sec32}) with multipliers
$$\mu_\pm(\lambda)=(N_1^\pm(\lambda))^2 \quad \mbox{and} \quad \nu_\pm(\lambda)=N_1^\pm (\lambda)N_2^\pm (\lambda)N_3^\pm (\lambda). $$
However, the calculations involved are rather big and we haven't found a neat way to present the operators.

\subsection{The Belov-Chaltikian lattice}
\begin{itemize}
\item Equation \cite{Belov93}:
\begin{eqnarray*}
\left\{ {\begin{array}{l} u_t= u (v_2-v_{-1})\\ v_t= u_{-1}-u + v (v_1-v_{-1}) \end{array} } \right.
\end{eqnarray*}
\item Hamiltonian structure \cite{Belov93}:
\begin{eqnarray*}
&&\cH_1=\begin{pmatrix}  u (\cE-\cE^{-1}) (\cE+1+\cE^{-1}) u &  u (\cE-1)(\cE+1+\cE^{-1}) v\\
v (1-\cE^{-1}) (\cE+1+\cE^{-1}) u &  v(\cE-\cE^{-1}) v +\cE^{-1} u-u\cE
\end{pmatrix},\qquad f_1=v\\
&&\cH_2=\begin{pmatrix}0
& u(1+\cE+\cE^2) \left( u \cE -\cE^{-2}u \right)\\
 \left(u \cE^{2} -\cE^{-1} u \right) (1+\cE^{-1}+\cE^{-2}) u & 
v(1+\cE) (u \cE-\cE^{-2} u)+(u\cE^2\!-\!\cE^{-1}\!u)(1+\cE^{-1})v
\end{pmatrix}\nonumber\\
&&\qquad+ \begin{pmatrix}\!  u (1+\cE+\cE^2)(\cE^{-1}\! v\!-\!v \cE)(1+\cE^{-1}\!+\cE^{-2}\!)u
& u (1+\cE+\cE^2)(\cE^{-1}\! v\!-\!v \cE)(1+\cE^{-1}\!)v\\
v (1+\cE) (\cE^{-1}\! v\!-\!v \cE)(1+\cE^{-1}\!+\cE^{-2}\!)u & v (1+\cE)(\cE^{-1}\! v\!-\!v \cE)(1+\cE^{-1}\!)v\! \end{pmatrix},\\
&&\qquad f_2=-\frac{1}{3} \ln u
\end{eqnarray*}
\item Non-trivial symmetry :
\begin{eqnarray*}
\cH_2 \delta f_1=\left( {\begin{array}{c}
 u v_{-1} \left( v+v_{-1}+v_{-2} \right) -u v_{2} \left( v_{1}+v_{2}+v_{3} \right) +u \left( u_{1}+u_{2}-u_{-1}-u_{-2} \right) 
\\
\left( u-vv_{1} \right)\left( v+v_{1}+v_{2} \right) + \left( vv_{-1}-u_{-1} \right)  \left( v+v_{-1}+v_{-2} \right) -v \left( u_{-2}-u_{1}\right)   
\end{array} } \right)
\end{eqnarray*}
\item Master symmetry \cite{sahadevan2}:
\begin{eqnarray*}
\left( \begin{array}{c}
n u_t+uv_{1}+4 u v_{-1}+u v   \\ n v_t +u-v v_{1}-4\,u_{-1}+4\,v v_{-1}+v^{2}
\end{array}  \right) 
\end{eqnarray*}
\item Lax representation \cite{hiin97}:
\begin{eqnarray*}
&&M= \left( \begin{array}{ccc} \lambda & \lambda v& \lambda u_{-1} \\ 1 & 0& 0\\0 & 1& 0\end{array}\right); \qquad
U= \left( \begin{array}{ccc} v-\lambda & -\lambda v & -\lambda u_{-1}\\ -1 & v_{-1} & 0\\0 & -1 & v_{-2}
\end{array}\right)
\end{eqnarray*}
\end{itemize}
The Belov-Chaltikian lattice is the Boussinesq lattice related to the
lattice $W_3$-algebra \cite{hiin97}. 

In recent paper \cite{beffawang12}, the authors wrote down the 
Boussinesq lattice related to the lattice $W_m$-algebra for the dependent variables
$u^1, u^2, \cdots, u^m$ and independent variables $n$ and $t$ as follows:
\begin{eqnarray*}
\left\{ \begin{array}{l}u^{1}_t=-u^{1} (u^{2}_m-u^{2}_{-1})\\ 
u^{i}_t=u^{i+1}-u^{i+1}_{-1}-u^{i} (u^{2}_{i-1}-u^{2}_{-1}), \quad i=2, 3\cdots, m-1\\
u^{m}_t=u^{1}-u^{1}_{-1}-u^{m} (u^{2}_{m-1}-u^{2}_{-1})
\end{array}\right.
\end{eqnarray*}
The vector ${\bf \tau}=(\tau^{1}, \cdots, \tau^{n})^{T}$ defined by
\begin{eqnarray*}
\begin{array}{l}\tau^{1}= n u^{1}_t -u^{1} \left((m+1) u^{2}_m+\sum_{l=0}^{m}u^{2}_{l}\right);\\ 
\tau^{i}= n u^{i}_t-u^{i} \left(i u^{2}_{i-1}+\sum_{l=0}^{i-1} u^{2}_{l}\right) +(i+1) u^{i+1}, \ \  i=2, 3\cdots, m-1\\
\tau^{m}= n u^{n}_t-u^{m} \left(m u^{2}_{m-1}+\sum_{l=0}^{m-1} u^{2}_{l}\right) +(m+1) u^{1}
\end{array}
\end{eqnarray*}
is a master symmetry. Its Hamiltonian structures are also studied in the same paper. 

\subsection{The Blaszak-Marciniak lattice}
\begin{itemize}
\item Equation \cite{BM94}:
\begin{eqnarray*}
\left\{ {\begin{array}{l} u_t= w_{1}-w_{-1}\\ v_t= u_{-1}w_{-1}-u w \\ w_t=  w(v-v_{1}) \end{array} } \right.
\end{eqnarray*}
\item Hamiltonian structure \cite{BM94}:
\begin{eqnarray*}
&&\cH_1=\!\!\left( {\begin{array}{ccc}
 \cE-\cE^{-1}&0&0  \\
  0&0&(\cE^{-1}-1)w\\
  0&-w(\cE-1)&0\\
 \end{array} } \right),\qquad f_1=uw+\frac{1}{2}v^2\\
&&\cH_2=\!\!\left(\!\!\! {\begin{array}{ccc}
\cE v \!-\! v \cE^{-1} \! \!-\!u (\cE+1)^{-1} (1\!-\!\cE) u & \cE w \cE\! -\!\cE^{-1} w & u (\cE+1)^{-1} (1-\cE) w \\
w \cE-\cE^{-1} w \cE^{-1} &  \cE^{-1} uw\! - \!uw \cE & v (\cE^{-1}-1) w\\
w(\cE+1)^{-1} (1-\cE) u &-w(\cE-1) v & w(\cE^{-1}\!-\!\cE) w \!-\! w (\cE+1)^{-1} (1\!-\!\cE) w
\end{array} }\!\!\! \right),\\&&\qquad f_2=v
\end{eqnarray*}
\item Recursion operator:
$$\cR=\cH_2 \cH_1^{-1}, $$
where
\begin{eqnarray*}
&&\cH_1^{-1}=\!\!\left( {\begin{array}{ccc}
 \frac{1}{2}  (\cE-1)^{-1}+\frac{1}{2} (\cE+1)^{-1}&0&0  \\
  0&0&- (\cE-1)^{-1} \frac{1}{w} \\
  0&-\frac{1}{w} \cE (\cE-1)^{-1}&0
 \end{array} } \right)
 \end{eqnarray*}
\item Non-trivial symmetry :
\begin{eqnarray*}
\cH_2 \delta f_1=\cH_2\left( {\begin{array}{c} w\\ v\\u
 \end{array} } \right)=\left( {\begin{array}{c}
w_1(v_1+v_2)-w_{-1}(v+v_{-1})\\
u_{-1}w_{-1}(v+v_{-1})-uw(v+v_1)-w_{-1}w_{-2}+ww_1\\
    w(v^2-v_1^2)+w(w_{-1}u_{-1}-w_1u_1)
 \end{array} } \right)
\end{eqnarray*}
\item Master symmetry: 
\begin{eqnarray*}
\cR \left(\begin{array}{c}  u/2 \\ v \\ 3 w/2 \end{array}\right)
\end{eqnarray*}
\item Lax representation \cite{BM94}:
\begin{eqnarray*}
L=\cS^2+u_1 \cS-v_1+w\cS^{-1}, \qquad A=\cS^2+u_1 \cS-v_1
\end{eqnarray*}
\end{itemize}
We do not explicitly write out its recursion operator, which is no longer weakly nonlocal although both operators $\cH_2$ and $\cH_1^{-1}$ are weakly nonlocal.
The statement that such recursion operator generates local symmetries can be proved in the same way as in \cite{wang09},
for weakly nonlocal differential recursion operators. We can compute the next Hamiltonian is $u v w +u v_1 w +\frac{v^3}{3}- w w_1$.
Its master symmetry is highly nonlocal, which is explicitly given in  \cite{sahadevan2}.

Another three-component lattice was given in \cite{zhang02} as
\begin{eqnarray*}
\left\{ {\begin{array}{l} p_t= q_{1} - q \\ q_t= q (p_{-1}-p)+r -r_{-1} \\ r_t=  r (p_{-1}-p_{1}) \end{array} } \right. .
\end{eqnarray*}
It has the Lax representation 
\begin{eqnarray*}
&&M= \left( \begin{array}{ccc} 0 & 1 & 0 \\ q & u+\lambda & 1\\r & 0& 0\end{array}\right); \qquad
U= \left( \begin{array}{ccc} -p_{-1} & 1 & 0\\ q & \lambda & 1\\r_{-1} & 0 & -p
\end{array}\right).
\end{eqnarray*}
This lattice is related to the Blaszak-Marciniak lattice by the Miura transformation 
$$p=v_1,\quad q=-u w, \quad r=-w w_1.$$
\section{Conclusion}
In this paper, we review two close concepts directly related to the Lax representations: Darboux transformations and Recursion operators
for integrable systems. We use the well-known nonlinear Schr{\"o}dinger equation, whose Lax representation is polynomial in the spectral parameter,
and a deformation of the derivative nonlinear Schr\"odinger equation, whose Lax representation is invariant 
under the dihedral reduction group $\bbbd_2$, as two typical examples. We then present a list of integrable differential-difference equations
containing equations themselves, Hamiltonian structures, recursion operators, a nontrivial generalised symmetries and Lax representations.
For most equations, we also add notes on their relations with other known equations and the weakly nonlocal inverse recursion operators if existing.

The theory of integrable partial difference (or discrete) equations is a relatively recent,
but very active area of research. There are famous the ABS equations for affine-linear quadrilateral
equations due to  Adler, Bobenko and Suris \cite{ABS, ABS1}, based on the condition of 3-D consistency.
Then Levi and Yamilov proposed to use  the existence of a generalized symmetry as a criteria for integrability
\cite{LeviYami2011} to classify integrable partial difference equations. They extended the ABS list.
For all integrable partial difference  equations, their symmetry flows can be viewed as integrable differential-difference
equations. The concept of recursion operator has been
adjusted to the difference equations to show that it generates an infinite
sequence of symmetries and canonical conservation laws for a partial difference equation \cite{mwx1,mwx2, MW2011}.
It can be proven that the difference equation shares the same recursion operator for its symmetry flows.
Thus the study the symmetry structure for integrable difference equations is in fact the same as for integrable differential-difference equations. 
This is one of the motivations to produce the list presented in this paper.
Besides, this list can serve as a benchmark for developing computer software packages  
for the symbolic computation of recursion operators, Lax representations, symmetries and conservation laws for nonlinear differential-difference equations \cite{hereman2}.

We have not presented here a rigorous proof that the operators we computed from  Darboux transformations are really Nijenhuis recursion operators for the
equations obtained from the corresponding Lax representations. When the Lax representations are polynomial in the spectral parameter and under certain technical conditions
a sketch of the  
proof was given in \cite{zhang02}. We think that the main statement of   \cite{zhang02} can be greatly generalised and simplified and there is a neat rigorous 
algebraic proof that the operators obtained from Darboux transformations invariant under reduction groups are indeed Nijenhuis recursion operators.

\section*{Acknowledgements}
The paper is supported by AVM's EPSRC grant EP/I038675/1 and JPW's EPSRC grant EP/I038659/1. Both authors gratefully acknowledge the financial support.

\end{document}